\documentclass[aip,reprint,nofootinbib]{revtex4-1}
\usepackage{graphicx,amsfonts,amsmath,amssymb}
\usepackage{bm}

\begin{document}
\title{Old Game, New Rules: Rethinking The Form of Physics}
\date{\today}
\author{C. Baumgarten}
\affiliation{5244 Birrhard, Switzerland}
\email{christian-baumgarten@gmx.net}

\def\begeq{\begin{equation}}
\def\endeq{\end{equation}}
\def\begary{\begeq\begin{array}}
\def\endary{\end{array}\endeq}
\def\bmtx{\left(\begin{array}}
\def\emtx{\end{array}\right)}
\def\eps{\varepsilon}
\def\d{\partial}
\def\y{\gamma}
\def\w{\omega}
\def\W{\Omega}
\def\s{\sigma}
\def\ket#1{\left|\,#1\,\right>}
\def\bra#1{\left<\,#1\,\right|}
\def\bracket#1#2{\left<\,#1\,\vert\,#2\,\right>}
\def\erw#1{\left<\,#1\,\right>}
\def\leftD#1{\overset{\leftarrow}{#1}}
\def\rightD#1{\overset{\rightarrow}{#1}}

\def\Exp#1{\exp\left(#1\right)}
\def\Log#1{\ln\left(#1\right)}
\def\Sinh#1{\sinh\left(#1\right)}
\def\Sin#1{\sin\left(#1\right)}
\def\Tanh#1{\tanh\left(#1\right)}
\def\Tan#1{\tan\left(#1\right)}
\def\Cos#1{\cos\left(#1\right)}
\def\Cosh#1{\cosh\left(#1\right)}

\begin{abstract}
We investigate the modeling capabilities of sets of coupled {\it classical harmonic
 oscillators} (CHO) in the form of a modeling game. The application of simple but restrictive 
rules of the game lead to conditions for an isomorphism between Lie-algebras and real
Clifford algebras. We show that the correlations between two coupled classical oscillators
find their natural description in the Dirac algebra and allow to model aspects of special
relativity, inertial motion, electromagnetism and quantum phenomena including spin in one go.
The algebraic properties of Hamiltonian motion of low-dimensional systems can generally
be related to certain types of interactions and hence to the dimensionality of emergent 
space-times. We describe the intrinsic connection between phase space volumes
of a 2-dimensional oscillator and the Dirac algebra. In this version of a
phase space interpretation of quantum mechanics the (components of the) spinor 
wavefunction in momentum space are abstract canonical coordinates, and the integrals 
over the squared wave function represents second moments in phase space. 
The wave function in ordinary space-time can be obtained via Fourier transformation. 
Within this modeling game, 3+1-dimensional space-time is interpreted as a structural 
property of electromagnetic interaction. A generalization selects a series of Clifford
algebras of specific dimensions with similar properties, specifically also 10- and 
26-dimensional real Clifford algebras.
\end{abstract}
\pacs{45.20.Jj, 47.10.Df, 41.75, 41.85, 03.65.Pm, 05.45.Xt, 03.30.+p, 03.65.-w,29.27.-a}
\keywords{Hamiltonian mechanics, Coupled Oscillators, Lorentz transformation, Dirac equation}
\maketitle

\section{Introduction}

D. Hestenes had the joyful idea to describe physics as a modeling game~\cite{Modeling0}.
We intend to play a modeling game with (ensembles of) classical harmonic 
oscillators (CHO). The CHO is certainly one of the most discussed and analyzed 
systems in physics and one of the few exactly solveable problems. 
One would not expect any substantially new discoveries related to this 
subject. Nevertheless there are aspects that are less well-known than others. 
One of these aspects concerns the transformation group of the symplectic 
transformations of $n$ coupled oscillators, $Sp(2n)$. 
We invite the reader to join us playing ``a modeling game'' and to discover
some fascinating features related to possible reinterpretations of systems
of two (or more) coupled oscillators. We will show that special relativity
can be reinterpreted as a transformation theory of the second moments of the
abstract canonical variables of coupled oscillator systems~\footnote{
The connection of the Dirac matrices to the symplectic group has been mentioned
by Dirac in Ref.~\cite{Dirac63}. For the connection of oscillators and Lorentz
transformations (LTs) see also the papers of Kim and Noz~\cite{KimNoz1981,KimNoz1982,KimNoz2003} 
and references therein. The use of CHOs to model quantum systems has been recently
described - for instance - by Briggs and Eisfeld~\cite{BriggEis}.}.
We extend the application beyond pure LTs and show that the Lorentz force 
can be reinterpreted by the second moments of two coupled oscillators in proper time. 
Lorentz transformations can be modeled as symplectic transformations~\cite{KimNoz1982}.
We shall show how Maxwell's equations find their place within the game.

The motivation for this game is to show that many aspects of modern physics
can be understood on the basis of the classical notions of harmonic oscillation
if these notions are appropriately reinterpreted.
 
In Sec.~\ref{sec_rules} we introduce the rules of our game, in 
Sec.~\ref{sec_osc} we introduce the algebraic notions of the 
Hamilton formalism. In Sec.~\ref{sec_geometry} we describe how geometry
emerges from coupled oscillator systems, in Sec.~\ref{sec_decoupling}
we describe the use of symplectic transformations and introduce
the Pauli- and Dirac algebra. In Sec.~\ref{sec_emeq} we introduce
a physical interpretation of oscillator moments and in Sec.~\ref{sec_mass}
we relate the phase space of coupled oscillators to the real Dirac algebra.
Sec.~\ref{sec_summary} contains a short summary.

\section{The Rules Of The Game}
\label{sec_rules}

The first rule of our game is the principle of reason (POR): 
{\it No distinction without reason} - we should not add or remove something 
{\it specific} (an asymmetry, a concept, a distinction) from our model without 
having a clear and explicite reason. If there is no reason for a specific
asymmetry or choice, then all possibilities are considered equivalently.
The second rule is the principle of variation (POV): We postulate that change is 
immanent to all fundamental quantities in our game. From these two rules, we 
take that the mathematical object of our theory is a list (n-tuple) of 
quantities (variables) $\psi$, each of which varies at all times. The third
rule is the principle of {\it objectivity} (POO): Any law within this game 
refers to measurements, defined as comparison of quantities (object
properties) in relation to other object properties of the same type (i.e. unit).
Measurements require rulers to enable for measurements. A measurement standard 
(ruler) has to be objective, i.e. based on properties of the objects of the
game. This apparent self-reference is unavoidable, as it models the {\it real}
situation of physics as experimental science.
Since all fundamental objects (quantities) in our model 
{\it vary at all times}, the only option to construct a constant quantity that
might serve as a ruler, is given by {\it constants of motion} (COM). Hence the 
principle of objectivity means that measurement standards are based on 
constants of motion.

This third rule implies that the fundamental variables can not be directly
measured, but only functions of the fundemantal variables of the same 
dimension (unit) of a COM. Thus the model has two levels: The level of the
fundamental variable list $\psi$, which is experimentally not directly 
accessible and a level of {\it observables} which are (as we shall argue) 
even moments of fundamental variables.

\subsection{Discussion of the Rules}
\label{sec_philo}

E.T. Jaynes wrote that 
``Because of their empirical origins, QM and QED are not physical theories 
at all. In contrast, Newtonian celestial mechanics, 
Relativity, and Mendelian genetics are physical theories, because their
mathematics was developed by reasoning out the consequences of clearly
stated physical principles from which constraint the possibilities''.
And he continues ``To this day we have no constraining principle from which
one can deduce the mathematics of QM and QED; [...] In other words, the 
mathematical system of the present quantum theory is [...] unconstrained by
any physical principle''~\cite{Jaynes}. This remarkably harsh criticism
of quantum mechanics raises the question of what we consider to be a
physical principle. Are the rules of our game physical principles?
We believe that they are no substantial physical principles but formal
first principles, they are {\it preconditions} of a sensible theory. 
They contain no immediate physical content, but they define the {\it form} 
or the {\it idea} of physics. 

It is to a large degree immanent to science and specifically to physics to 
presuppose the existence of {\it reason}: Apples do not fall down by chance - 
there is a reason for this tendency. Usually this believe in reason implies 
the believe in causality, i.e. that 
we can also (at least in principle) explain why a specific apple falls at a
specific time, but practically this latter believe can rarely be confirmed
experimentally and therefore remains to some degree metaphysical.
Thus, if, as scientists, we postulate that things have reason, then this
is no {\it physical} principle but a precondition, a first principle.

The second rules (POV), is specific to the form (or idea) of physics, 
e.g. that it is the {\it sense} of physics to {\it recognize the pattern}
of motion and to {\it predict future}. Therefore the notion of time in
the form of change is indeed immanent to any physical description of reality.
The principle of objectivity is immanent to the very idea of physics:
A measurement is the comparison of properties of objects with compatible 
properties of reference objects, e.g. requires ``constant'' rulers. 
Hence the rules of the game are to a large degree unavoidable: They follow 
from the very form of physics and therefore certain laws of physics are
not substantial results of a physical theory. For instance a consistent
``explanation'' of the stability of matter is impossible {\it as we presumed 
it already within the idea of measurement}. More precisely: if this presumption
does {\it not} follow within the framework of a physical theory, then the 
theory is fundamentally flawed, since it can not even reproduce it's own
presumptions.

Einstein wrote with respect to relativity that
``It is striking that the theory (except for the four-dimensional space)
introduces two kinds of things, i.e. (1) measuring rods and clocks, (2)
all other things, e.g., the electromagnetic field, the material point, etc.
This, in a certain sense, is inconsistent; strictly speaking, measuring rods
and clocks should emerge as solutions of the basic equations [...], not, as
it were, as theoretically self-sufficient entities.''~\cite{EinsteinBio}.
The more it may surprise that the stability of matter can not be obtained
from classical physics as remarked by Elliott H. Lieb: ``A fundamental paradox
of classical physics is why matter, which is held together by Coulomb forces,
does not collapse''~\cite{Lieb}. This single sentence seems to rule out the 
possibility of a fundamental classical theory and uncovers the uncomfortable
situation of theoretical physics today: Despite the overwhelming experimental
and technological success, there is a deep-seated confusion concerning the 
theoretical foundations. Our game is therefore a meta-experiment. It is not
the primary goal to find ``new'' laws of nature or new experimental 
predictions, but it is a conceptional ``experiment'' that aims to further 
develop our understanding of the consequences of principles: which ones
are {\it really} required to derive central ``results'' of contemporary 
physics. In this short essay final answers can not be given, but maybe
some new insights are possible.

\subsection{What about Space-Time?}
\label{sec_spacetime}

A theory has to make the choice between postulate and proof. 
If a 3+1 dimensional space-time is presumed, then it cannot be proven within 
the same theoretical framework. Or at least the value of such proof remains 
questionable. This is a sufficient
reason to refuse any postulates concerning the dimensionality of space-time.
Another - even stronger - reason to avoid a direct postulate of space-time
and its geometry has been given above: The fundamental variables that we 
postulated above, can not be directly measured. This excludes space-time 
coordinates as primary variables (which can be directly measured), but with 
it almost all other apriori assumed concepts like velocity, acceleration, 
momentum, energy and so on. At some point these concepts certainly have to 
be introduced, but we suggest an approach to the formation of concepts that 
differs from the Newtonian axiomatic method. The POR does not allow to introduce
distinctions between the fundamental variables into coordinates and momenta
without reason. Therefore we are forced to use an interpretational method,
which one might summarize as {\it function follows form}. We shall first derive
equations and then we shall interpret the equations according to some formal
criteria. This implies that we have to refer to already existing notions
if we want to identify quantities according to their appearance within a 
certain formalism. The consequence for the game is, that we have to show 
how to give rise to {\it geometrical} notions: If we do not postulate
space-time then we apparently have to construct it.

A consequence of our conception is that both, objects and fields have to be 
identified with dynamical structures, as there is simply nothing else
available. This nicely fits to the framework of structure preserving
(symplectic) dynamics that follows from the described principles.

\section{Theory of Small Oscillations}
\label{sec_osc}

In this section we shall derive the theory of coupled oscillators from the rules
of our game. According to the POO there exists a function (COM) ${\cal
  H}(\psi)$ such that~\footnote{
Let us first (for simplicity) assume that ${\d{\cal H}\over\d t}=0$.}:
\begeq
{d{\cal H}\over dt}=\sum\limits_k\,{\d{\cal H}\over\d\psi_k}\,\dot\psi_k=0\,,
\endeq
or in vector notation 
\begeq
{d{\cal H}\over dt}=(\nabla_\psi\,{\cal H})\,\cdot\dot\psi=0\,.
\label{eq_heqom0}
\endeq
The simplest solution is given by an arbitrary skew-symmetric matrix ${\cal X}$:
\begeq
\dot\psi={\cal X}\,\nabla_\psi\,{\cal H}\,.
\label{eq_heqom}
\endeq
Note that it is only the {\it skew-symmetry} of ${\cal X}$, which ensures
that it is always a solution to Eqn.~\ref{eq_heqom0} and which ensures 
that ${\cal H}$ is constant. 
If we now consider a state vector $\psi$ of dimension $k$, then there 
is a theorem in linear algebra, which states that for {\it any} skew-symmetric 
matrix ${\cal X}$ there exists a non-singular matrix ${\bf Q}$ such that 
we can write~\cite{MHO1}:
\begeq
{\bf Q}^T\,{\cal X}\,{\bf Q}=\textrm{diag}(\eta_0,\eta_0,\eta_0,\dots\,,0,0,0)\,.
\label{eq_strucdef}
\endeq
where $\eta_0$ is the matrix
\begeq
\eta_0=\bmtx{cc}
0&1\\
-1&0\\
\emtx\,.
\label{eq_eta0}
\endeq
If we restrict us to orthogonal matrices ${\bf Q}$, then we may still write
\begeq
{\bf Q}^T\,{\cal X}\,{\bf Q}=\textrm{diag}(\lambda_0\,\eta_0,\lambda_1\,\eta_0,\lambda_2\,\eta_0,\dots\,,0,0,0)\,.
\label{eq_strucdeforth}
\endeq
In both cases we may leave away the zeros, since they correspond to
non-varying variable, which is in conflict with the second rule 
of our modeling game. Hence $k=2n$ must be even and the square matrix 
${\cal X}$ has the dimension $2n\times 2n$. As we have no specific reason 
to assume asymmetries between the different degrees of freedom (DOF), we have 
to choose all $\lambda_k=1$ in Eqn.~\ref{eq_strucdeforth} and return to 
Eqn.~\ref{eq_strucdef} without zeros and define the block-diagonal
so-called {\it symplectic unit matrix} (SUM) $\y_0$:
\begeq
{\bf Q}^T\,{\cal X}\,{\bf Q}=\textrm{diag}(\eta_0,\eta_0,\dots,\eta_0)\equiv\y_0\,.
\label{eq_strucdef0}
\endeq
These few basic rules thus lead us directly to Hamiltonian
mechanics: Since the state vector has even dimension and due to
the form of $\y_0$, we can interpret $\psi$ as an ensemble of $n$ 
classical DOF - each DOF represented by a canonical pair 
of coordinate and momentum: $\psi=(q_1,p_1,q_2,p_2,\dots\,,q_n,p_n)^T$. 
In this notation and after the application of the transformation ${\bf Q}$, 
Eqn.~\ref{eq_heqom} can be written in form of the Hamiltonian equations of
motion (HEQOM):
\begary{rcl}
\dot q_i&=&{\d{\cal H}\over \d p_i}\\
\dot p_i&=&-{\d{\cal H}\over \d q_i}\\
\label{eq_heqom2}
\endary
The validity of the HEQOM is of fundamental importance as it allows for the
use of the results of Hamiltonian mechanics, of statistical mechanics 
and thermodynamics - but without the intrinsic presupposition that the 
$q_i$ have to be understood as positions in real space and the $p_i$ as 
the corresponding canonical momenta. This is legitimate as the theory of 
canonical transformations is {\it independent from 
any specific physical interpretation of what the coordinates and momenta 
represent physically}.
The canonical pairs are coordinates $q_i, p_i$ in an abstract phase space 
and they can be interpreted as canonical coordinates and momenta due to 
the {\it form} of the HEQOM. The choice of the specific form of $\y_0$ is for $n>1$
DOF not unique. It could for instance be written as 
\begeq
\y_0\equiv\eta_0\,\otimes\,{\bf 1}_{n\times n}\,,
\endeq
which corresponds a state vector of the form
$$\psi=(q_1,\dots,q_n,p_1,\dots,p_n,)^T\,,$$ 
or by 
\begeq
\y_0\equiv{\bf 1}_{n\times n}\,\otimes\eta_0\,,
\endeq
as in Eq.~\ref{eq_strucdef0}. Therefore we are forced to make an arbitrary 
choice~\footnote{But we should keep in mind, that other ``systems'' with a different
choice are possible. If we can not exclude their existence, then they should
exist as well. With respect to the form of the SUM, we suggest that different
``particle'' types (different types of fermions for instance) have a different 
SUM.}. But in all cases the SUM $\y_0$ must be
skew-symmetric and have the following properties:
\begary{rcl}
\y_0^T&=&-\y_0\\
\y_0^2&=&-{\bf 1}\,,
\endary
which also implies that $\y_0$ is orthogonal and has unit determinant. 
Note also that all eigenvalues of $\y_0$ are purely imaginary.
However, once we have chosen a specific form of $\y_0$, we have
specified a set of canonical pairs $(q_i,p_i)$ within the state vector.
This choice fixes the set of possible canonical (structure preserving)
transformations.

Now we write the Hamiltonian ${\cal H}(\psi)$ as a Taylor series, we remove the 
rule-violating constant term and cut it after the second term. We do not
claim that higher terms may not appear, but we delay the discussion of higher
orders to a later stage. All this is well-known in the theory of small oscillations.
There is only one difference to the conventional treatment: We have no direct
macroscopic interpretation for $\psi$ and following our first rule we have to 
write the second-order Hamiltonian ${\cal H}(\psi)$ in the most general form:
\begeq
{\cal H}(\psi)=\frac{1}{2}\,\psi^T\,{\cal A}\,\psi\,,
\label{eq_cho_hamiltonian}
\endeq
where ${\cal A}$ is only restricted to be {\it symmetric} as all non-symmetric
terms {\it do not contribute} to ${\cal H}$. 
Since it is not unlikely to find more than a single constant of motion 
in systems with multiple DOFs, we distinguish systems with singular matrix ${\cal A}$
from those with a positive or negative definite matrix ${\cal A}$. 
Positive definite matrices are favoured in the sense that they allow 
to identify ${\cal H}$ with the amount of a substance or an amount of
energy~\footnote{It is immanent to the concept of substance 
that it is understood as something positive semidefinite.}.

Before we try to interprete the elements in ${\cal A}$, we will
explore some general algebraic properties of the Hamiltonian formalism. 
If we plug Eqn.~\ref{eq_cho_hamiltonian} into Eqn.~\ref{eq_heqom}, then
the equations of motion can be written in the general form
\begeq
\dot\psi=\y_0\,{\cal A}\,\psi={\bf F}\,\psi\,.
\label{eq_eqom_cho}
\endeq
The matrix ${\bf F}=\y_0\,{\cal A}$ is the product of the symmetric (positive
semi-definite) matrix ${\cal A}$ and the skew-symmetric matrix $\y_0$. 
As known from linear algebra, the trace of such products is zero:
\begeq
\mathrm{Tr}({\bf F})=0\,.
\endeq
It is obvious that pure harmonic oscillation of $\psi$ is given for matrices 
${\bf F}$ that have purely imaginary eigenvalues. Furthermore these are the only
stable solutions~\cite{MHO}. Note that Eq.~\ref{eq_eqom_cho} may represent a
tremendous amount of different types of systems - all linearily coupled 
systems in any dimension, chains or $d$-dimensional lattices of linear coupled 
oscillators and wave propagation~\footnote{However the linear approximation does
not allow for the description of the transport of heat.}.

One quickly derives from the properties of $\y_0$ and ${\cal A}$ that 
\begeq
{\bf F}^T={\cal A}^T\,\y_0^T=-{\cal A}\,\y_0=\y_0^2\,{\cal A}\,\y_0=\y_0\,{\bf F}\,\y_0\,.
\label{eq_symplex}
\endeq
Since any square matrix can be written as the sum of a symmetric 
and a skew-symmetric matrix, it is nearby to also consider the properties
of products of $\y_0$ with a skew-symmetric real square matrices ${\cal B}$.
If ${\bf C}=\y_0\,{\cal B}$, then
\begeq
{\bf C}^T={\cal B}^T\,\y_0^T={\cal B}\,\y_0=-\y_0^2\,{\cal B}\,\y_0=-\y_0\,{\bf C}\,\y_0\,.
\label{eq_cosymplex}
\endeq
Symmetric $2n\times 2n$-matrices contain $2n\,(2n+1)/2$ different matrix
elements and skew-symmetric ones $2n\,(2n-1)/2$ elements, so that there
are $\nu_s$ linear independent symplices 
\begeq
\nu_s=n\,(2n+1)
\label{eq_ns}
\endeq 
and $\nu_c$ cosymplices with 
\begeq
\nu_c=n\,(2n-1)\,.
\label{eq_nc}
\endeq 
In the theory of linear Hamiltonian dynamics, matrices of the form of ${\bf
  F}$ are known as ``Hamiltonian'' or ``infinitesimal symplectic'' and those
of the form of ${\bf C}$ as ``skew-Hamiltonian'' matrices. 
This convention is a bit odd as ${\bf F}$ does not appear in the Hamiltonian 
and it is in general not symplectic. Furthermore the term ``Hamiltonian matrix''
has a different meaning in quantum mechanics - more in analogy to ${\cal A}$.
But it is known that this type of matrix is closely connected to symplectic matrices 
as every symplectic matrix is a matrix exponential of a matrix ${\bf F}$~\cite{MHO}.
We consider the matrices as defined by Eqn.~\ref{eq_symplex} and Eqn.~\ref{eq_cosymplex} 
as too important and fundamental to have no meaningful and unique names:
Therefore we speak of a {\bf symplex} (plural {\it symplices}), if a matrix holds 
Eqn.~\ref{eq_symplex} and of a {\bf cosymplex} if it holds Eqn.~\ref{eq_cosymplex}.

\subsection{Symplectic Motion and Second Moments}

So what is a symplectic matrix anyway? The concept of symplectic transformations
is a specific formulation of the theory of canonical transformations.
Consider we define a new state vector (or new coordinates) $\phi(\psi)$ - with
the additional requirement, that the transformation is reversible. Then
the Jacobian matrix of the transformation is given by
\begeq
{\bf J}_{ij}=\left({\d\phi_i\over\d\psi_j}\right)\,,
\endeq
and the transformation is said to be symplectic, if the Jacobian matrix holds~\cite{MHO}
\begeq
{\bf J}\,\y_0\,{\bf J}^T=\y_0\,.
\label{eq_symplectic}
\endeq
Let us see what this implies in the linear case:
\begary{rcl}
{\bf J}\,\dot\psi&=&{\bf J}\,{\bf F}\,{\bf J}^{-1}\,{\bf J}\,\psi\\
\tilde\psi&=&{\bf J}\,\psi\\
\dot{\tilde\psi}&=&{\bf J}\,{\bf F}\,{\bf J}^{-1}\,\tilde\psi\\
\dot{\tilde\psi}&=&{\bf\tilde F}\,\tilde\psi\\
\endary
and - by the use of Eqn.~\ref{eq_symplectic} one finds that ${\bf\tilde F}$ is
still a symplex:
\begary{rcl}
{\bf\tilde F}^T&=&({\bf J}^{-1})^T\,{\bf F}^T\,{\bf J}^T\\
{\bf\tilde F}^T&=&({\bf J}^{-1})^T\,\y_0\,{\bf F}\,\y_0\,{\bf J}^T\\
{\bf\tilde F}^T&=&-\y_0^2\,({\bf J}^{-1})^T\,\y_0\,{\bf F}\,{\bf J}^{-1}\,\y_0\\
{\bf\tilde F}^T&=&-\y_0\,{\bf J}\,\y_0^2\,{\bf F}\,{\bf J}^{-1}\,\y_0\\
{\bf\tilde F}^T&=&\y_0\,{\bf J}\,{\bf F}\,{\bf J}^{-1}\,\y_0\\
{\bf\tilde F}^T&=&\y_0\,{\bf\tilde F}\,\y_0\\
\endary
Hence a symplectic transformation is first of all a similarity transformation, but
secondly, it preserves the structure of all involved equations. Therefore the
transformation is said to be {\it canonical} or {\it structure preserving}. 
The distinction between canonical and non-canonical transformations can therefore 
be traced back to the skew-symmetry of $\y_0$ and the symmetry of ${\cal A}$ - 
both of them consequences of the rules of our physics modeling game.

Recall that we argued that the matrix ${\cal A}$ should be symmetric {\it
  because} skew-symmetric terms do not contribute to the Hamiltonian. 
Let us have a closer look what this means. Consider the matrix of second
moments $\Sigma$ that can be build from the variables $\psi$:
\begeq
\Sigma\equiv\langle\psi\,\psi^T\rangle\,,
\endeq
in which the angles indicate some (yet unspecified) sort of average. The
equation of motion of this matrix is given by
\begary{rcl}
\dot\Sigma&=&\langle\dot\psi\,\psi^T\rangle+\langle\psi\,\dot\psi^T\rangle\\
\dot\Sigma&=&\langle{\bf F}\,\psi\,\psi^T\rangle+\langle\psi\,\psi^T\,{\bf F}^T\rangle\,.
\endary
Now, as long as ${\bf F}$ does not depend on $\psi$, we obtain
\begary{rcl}
\dot\Sigma&=&{\bf F}\,\Sigma+\Sigma\,{\bf F}^T\\
\dot\Sigma&=&{\bf F}\,\Sigma+\Sigma\,\y_0\,{\bf F}\,\y_0\\
(\dot\Sigma\,\y_0)&=&{\bf F}\,(\Sigma\,\y_0)-(\Sigma\,\y_0)\,{\bf F}\,,
\endary
so that with the matrix ${\bf S}\equiv\Sigma\,\y_0$ this yields:
\begary{rcl}
{\bf\dot S}&=&{\bf F}\,{\bf S}-{\bf S}\,{\bf F}\,.
\label{eq_envelope}
\endary
This equation is a key to our theory. Firstly it contains only observables: though
we started with unmeasurable abstract quantities, we reach here a stage in which
the original quantities $\psi$ are completely hidden. In accelerator physics this
equation is called {\it envelope equation} as it describes the dynamics of the 
so-called ``envelope'' of the beam, which is nothing but a parametrization by
statistical (here: second) moments.

For completeness we introduce the ``adjunct'' spinor $\bar\psi=\psi^T\,\y_0$ so that we may write
\begeq
{\bf S}=\langle\,\psi\bar\psi\,\rangle\,.
\endeq
Note that ${\bf S}$ is also a symplex. The matrix ${\bf S}$ (i.e. all second
moments) is constant, iff ${\bf S}$ and ${\bf F}$ commute. 

Now we define an {\it observable} to be an operator ${\bf O}$ with a (potentially) 
non-vanishing expectation value, defined by:
\begeq
\langle{\bf O}\rangle\equiv\langle\bar\psi{\bf O}\psi\rangle=\langle\psi^T\,\y_0\,{\bf O}\psi\rangle\,.
\endeq
Thus, if the product $\y_0\,{\bf O}$ is {\it not} skew-symmetric, i.e. contains 
a product of $\y_0$ with a symmetric matrix ${\cal B}$, then the expectation value 
is potentially non-zero: 
\begeq
\langle{\bf O}\rangle\equiv\langle\psi^T\,\y_0\,(\y_0\,{\cal B})\psi\rangle=-\langle\psi^T\,{\cal B}\,\psi\rangle\,.
\endeq
This means that only the symplex-part of an operator is ``observable'', while cosymplices
yield a vanishing expectation value. Hence Eq.~\ref{eq_envelope} delivers the blueprint 
for the general definition of observables. Furthermore we find in the last line the 
constituting equation for Lax pairs~\cite{Lax}. Peter Lax has shown that
for such pairs of operators ${\bf S}$ and ${\bf F}$ that obey Eqn.~\ref{eq_envelope} 
there are the following constants of motion
\begeq
\mathrm{Tr}({\bf S}^k)=\mathrm{const}\,.
\label{eq_laxconst}
\endeq
for arbitrary integer $k>0$. Since ${\bf S}$ is a symplex and therefore by 
definition the product of a symmetric matrix and the skew-symmetric $\y_0$, 
Eqn.~\ref{eq_laxconst} is always zero and hence trivially true for $k=1$. 
The same is true for any odd power of ${\bf S}$, as it can be easily shown
that any odd power of a symplex is again a symplex (see Eq.~\ref{eq_cosy_algebra}),
so that the only non-trivial general constants of motion correspond to even 
powers of ${\bf S}$, which implies that all observables are functions of even 
powers of the fundamental variables.

To see the validity for $k>1$ we have to consider
the general algebraic properties of the trace operator.
Let $\lambda$ be an arbitrary real constant and $\tau$ be a real parameter, then
\begary{rcl}
\mathrm{Tr}({\bf A})&=&\mathrm{Tr}({\bf A}^T)\\
\mathrm{Tr}(\lambda\,{\bf A})&=&\lambda\,\mathrm{Tr}({\bf A})\\
{d\over d\tau}\mathrm{Tr}({\bf A}(\tau))&=&\mathrm{Tr}({d{\bf A}\over d\tau})\\
\mathrm{Tr}({\bf A}+{\bf B})&=&\mathrm{Tr}({\bf A})+\mathrm{Tr}({\bf B})\\
\mathrm{Tr}({\bf A}\,{\bf B})&=&\mathrm{Tr}({\bf B}\,{\bf A})\,.
\endary
It follows that 
\begary{rcl}
0&=&\mathrm{Tr}({\bf A}\,{\bf B}-{\bf B}\,{\bf A})\\
0&=&\mathrm{Tr}({\bf A}^{n}\,{\bf B}-{\bf A}^{n-1}\,{\bf B}\,{\bf A})\\
0&=&\mathrm{Tr}\left[{\bf A}^{n-1}\,({\bf A}\,{\bf B}-{\bf B}\,{\bf A})\right]\\
\label{eq_lax0}
\endary
From the last line of Eqn.~\ref{eq_lax0} follows for ${d{\bf A}\over d\tau}=\lambda\,({\bf
  A}\,{\bf B}-{\bf B}\,{\bf A})$ 
\begeq
{d\over d\tau}\mathrm{Tr}({\bf A}^n)=0
\endeq
Remark: This conclusion is not limited to symplices.

However for single spinors $\psi$ and their second moments ${\bf
  S}=\Sigma\,\y_0=\psi\psi^T\,\y_0$ we find:
\begary{rcl}
\mathrm{Tr}({\bf S}^k)&=&\mathrm{Tr}[\psi\,\psi^T\,\y_0\cdots\psi\,\psi^T\,\y_0]\\
                      &=&\mathrm{Tr}[\psi\,(\psi^T\,\y_0\cdots\psi\,\psi^T\,\y_0)]\\
                      &=&\mathrm{Tr}[(\psi^T\,\y_0\cdots\psi\,\psi^T\,\y_0)\,\psi]\\
                      &=&\mathrm{Tr}\left[(\psi^T\,\y_0\,\psi)\cdots(\psi^T\y_0\psi)\right]=0\\
\endary
since each single factor $(\psi^T\,\y_0\,\psi)$ vanishes due to the
skew-symmetry of $\y_0$. Therefore the constants of motion as derived from
Eqn.~\ref{eq_laxconst} are non-zero only for even $k$ and {\it after averaging 
over some kind of distribution} such that ${\bf S}=\langle \psi\psi^T\,\y_0\rangle$
has non-zero eigenvalues as in Eq.~\ref{eq_composite} below.

The symmetric matrix $2n\times 2n$-matrix
$\Sigma$ (and also ${\cal A}$) is positive definite, if it can be written
as a pro\-duct $\Sigma=\Psi\Psi^T$, where $\Psi$ is a non-singular matrix
of size $2n\times m$ with $m\ge 2n$.

For $n=m/2=1$, the form of $\Psi$ may be chosen as 
\begary{rcl}
\Psi&=&{1\over\sqrt{q^2+p^2}}\,\bmtx{cc}
q&-p\\
p&q\\
\emtx={1\over\sqrt{q^2+p^2}}\,({\bf 1}\psi,\eta_0\,\psi)\\
\Rightarrow&&\Sigma=\Psi\Psi^T=\Psi^T\,\Psi={\bf 1}\\
{\bf S}&=&\y_0\,
\label{eq_composite}
\endary
so that for $k=2$ the average of two ``orthogonal'' column-vectors $\psi$ and
$\eta_0\,\psi$ gives a non-zero constant of motion via Lax pairs as
$\y_0^2=-{\bf 1}$. 

These findings have numerous consequences for the modeling game. 
The first is that we have found constants of motion - though some of them 
are physically meaningful only for a non-vanishing volume in phase space, 
i.e. by the combination of several spinors $\psi$. Secondly, a stable 
state ${\bf\dot S}=0$ implies that the matrix operators forming the 
Lax pair have the same eigenvectors: a density distribution in phase 
space (as described by the matrix of second moments) is stable if it is 
adapted or {\it matched} to the symplex ${\bf F}$. 
The phase space distribution as represented by ${\bf S}$ and the driving terms 
(the components of ${\bf F}$) must fit to each other in order to obtain a 
stable ``eigenstate''. But we also found a clear reason, why generators 
(of symplectic transformations) are always observables and vice versa:
Both, the generators as well as the observables are symplices of the same
type. There is a one-to-one correspondence between them, not only as
{\it generators of infinitesimal transformations}, but also algebraically. 

Furthermore, we may conclude that (anti-) commutators are an essential part of 
``classical'' Hamiltonian mechanics and secondly that the matrix ${\bf S}$ has 
the desired properties of observables: Though ${\bf S}$ is based on
continuously varying fundamental variables, it is constant, if it commutes 
with ${\bf F}$, and it varies otherwise~\footnote{In accelerator
physics, Eqn.~\ref{eq_envelope} describes the envelope of a beam in linear optics.
The matrix of second moments $\Sigma$ is a covariance matrix - and therefore our
modeling game is connected to probability theory exactly at the stage where we define
observables.}.

Hence it appears sensible to take a closer look on the (anti-) commutation relations
of (co-) symplices and though the definitions of (co-) symplices are quite plain, the 
(anti-) commutator algebra that emerges from them has a surprisingly rich structure. 
If we denote symplices by ${\bf S}_k$ and cosymplices by ${\bf C}_k$, then the 
following rules can quickly be derived:
\begary{ccc}
\left.\begin{array}{c}
{\bf S}_1\,{\bf S}_2-{\bf S}_2\,{\bf S}_1\\
{\bf C}_1\,{\bf C}_2-{\bf C}_2\,{\bf C}_1\\
{\bf C}\,{\bf S}+{\bf S}\,{\bf C}\\
{\bf S}^{2\,n+1}\\
\end{array}\right\} & \Rightarrow & \mathrm{symplex}\\&&\\
\left.\begin{array}{c}
{\bf S}_1\,{\bf S}_2+{\bf S}_2\,{\bf S}_1\\
{\bf C}_1\,{\bf C}_2+{\bf C}_2\,{\bf C}_1\\
{\bf C}\,{\bf S}-{\bf S}\,{\bf C}\\
{\bf S}^{2\,n}\\
{\bf C}^n\\
\end{array}\right\} & \Rightarrow & \mathrm{cosymplex}\\
\label{eq_cosy_algebra}
\endary
This {\it Hamiltonian} algebra of (anti-)commutators is of fundamental
importance insofar as we derived it in a few steps from first principles
(i.e. the rules of the game) and it defines the structure of Hamiltonian 
dynamics in phase space. The distinction between symplices and cosymplices 
is also the distinction between observables and non-observables. It is the
basis of essential parts of the following considerations.

\section{Geometry from Hamiltonian Motion}
\label{sec_geometry}

In the following we will demonstrate the {\it geometrical content} of the
algebra of (co-)symplices (Eqn.~\ref{eq_cosy_algebra}) which emerges
for {\it specific numbers of DOF $n$}. As shown above pairs
of canonical variables (DOFs) are the a direct consequence of the abstract
rules of our game. Though single DOFs are poor ``objects'', it is remarkable
to find physical structures emerging from our abstract rules {\it at all}. 
This suggests that there might be more structure to discover when $n$
DOF are combined, for instance geometrical structures. The following 
considerations obey the rules of our game, since they are based purely on 
symmetry considerations like those that guided us towards Hamiltonian dynamics.
The objects of interest in our algebraic interpretation of Hamitonian 
dynamics are matrices. The first matrix (besides ${\cal A}$) with a specific 
form that we found, is $\y_0$. It is a symplex:
\begeq
\y_0^T=-\y_0=\y_0\,\y_0\,\y_0\,.
\endeq
According to Eq.~\ref{eq_ns} there are $\nu_s=n\,(2\,n+1)$ (i.e. $\nu_s\ge 3$)
symplices. Hence it is nearby to ask if other symplices with similar 
properties like $\y_0$ exist - and if so, what the relations between these 
matrices are. According to Eq.~\ref{eq_cosy_algebra} the commutator of two 
symplices is again a symplex, while the anti-commutator is a cosymplex.
Hence - as we are interested in {\it observables} and components of the 
Hamiltonians (i.e. symplices), respectively, we would like to find other 
symplices that anti-commute with $\y_0$ and with each other. 
In this case, the product of two such matrices is also a symplex, i.e. 
another potential contribution to the general Hamiltonian matrix ${\bf F}$.

Assumed we had a set of $N$ mutually anti-commuting orthogonal symplices $\y_0$
and $\y_k$ with $k\in\,[1\dots N-1]$, then a Hamiltonian matrix ${\bf F}$
might look like
\begary{rcl}
{\bf F}=\sum\limits_{k=0}^{N-1}\,f_k\,\y_k+\dots\,.
\endary
The $\y_k$ are symplices {\it and} anti-commute with $\y_0$:
\begeq
\y_0\,\y_k+\y_k\,\y_0=0\,.
\endeq
Multiplication from the left with $\y_0$ gives:
\begeq
-\y_k+\y_0\,\y_k\,\y_0=-\y_k+\y_k^T=0\,.
\endeq
Hence all other possible symplices $\y_k$ that anticommute with $\y_0$ are symmetric -
and hence they square to ${\bf 1}$ as they are also assumed to be orthogonal.
This is an extremely important finding for what follows, as it can (within our
game) be interpreted as a classical proof of the uniqueness of (observable) 
time-dimension: Time is one-dimensional as there is no other skew-symmetric
symplex that anti-commutes with $\y_0$. We can choose different forms for
$\y_0$, but the emerging algebra allows for no second ``direction of time''.

The second order derivative of $\psi$ is (for constant ${\bf F}$) given by
$\ddot\psi={\bf F}^2\,\psi$ which yields:
\begary{rcl}
{\bf F}^2=\sum\limits_{i=0}^{N-1}\,f_i^2\,\y_i^2+\sum\limits_{i\ne j}\,f_i\,f_j\,(\y_i\,\y_j+\y_j\,\y_i)\,.
\endary
Since the matrices on the right anticommute, we are left with:
\begary{rcl}
{\bf F}^2=\left(\sum\limits_{k=1}^{N-1}\,f_k^2-f_0^2\right)\,{\bf 1}\,.
\label{eq_minkowski}
\endary
Thus - we find a set of (coupled) oscillators, if 
\begeq
f_0^2>\sum\limits_{k=1}^{N-1}\,f_k^2\,,
\endeq
such that 
\begeq
\ddot\psi=-\omega^2\,\psi\,.
\endeq
Given such matrix systems exist - then they generate a Minkowski type ``metric'' 
as in Eq.~\ref{eq_minkowski}~\footnote{
Indeed it appears that Dirac derived his system of matrices from the this 
requirement~\cite{DiracMatrices}.}. The appearance of such a metric may guide
us towards further aspects of physics to be modeled.
It should be possible to construct a Minkowski type geometry from the driving 
terms of oscillatory motion. This is indeed possible - 
at least for symplices of certain dimensions as we will show below.
The first thing needed is some kind of measure to define the length of a ``vector''. 
Since the length is a measure that is invariant under certain 
transformations, specifically under rotations, we prefer to use a quantity with 
certain invariance properties to define a length. The only one we have at hand is 
given by Eqn.~\ref{eq_laxconst}. Accordingly we define the (squared) length of a 
matrix representing a ``vector'' by
\begeq
\Vert {\bf A}\Vert^2\equiv\frac{1}{2\,n}\,\mathrm{Tr}({\bf A}^2)\,.
\label{eq_norm}
\endeq
The division by $2\,n$ is required to make the unit matrix have unit norm.
Besides the norm we need a scalar product, i.e. a definition of orthogonality. 
Consider the Pythagorean theorem which says that two vectors
$\vec a$ and $\vec b$ are orthogonal iff
\begeq
(\vec a+\vec b)^2=\vec a^2+\vec b^2\,.
\label{eq_p1}
\endeq
The general expression is
\begeq
(\vec a+\vec b)^2=\vec a^2+\vec b^2+2\,\vec a\cdot\vec b\,.
\label{eq_p2}
\endeq
The equations are equal, iff $\vec a\cdot\vec b=0$. Hence the Pythagorean
theorem yields a reasonable definition of orthogonality. However, we had
no method yet to define vectors within our game.
Using matrices ${\bf A}$ and ${\bf B}$ we may then write
\begary{rcl}
\Vert {\bf A}+{\bf B}\Vert^2&=&\frac{1}{2\,n}\,\mathrm{Tr}\left[({\bf A}+{\bf B})^2\right]\\
&=&\Vert {\bf A}\Vert^2+\Vert{\bf B}\Vert^2+\frac{1}{2\,n}\,\mathrm{Tr}({\bf A}{\bf B}+{\bf B}{\bf A})\,.
\endary
If we compare this to Eqn.~\ref{eq_p1} and Eqn.~\ref{eq_p2}, respectively,
then the obvious definition of the inner product can be defined as follows:
\begeq
{\bf A}\cdot{\bf B}\equiv{{\bf A}\,{\bf B}+{\bf B}\,{\bf A}\over 2}
\label{eq_inner}
\endeq
Since the anticommutator does in general not yield a scalar, we
have to distinguish between inner product and scalar product:
\begeq
({\bf A}\cdot{\bf B})_S\equiv\frac{1}{4\,n}\,\mathrm{Tr}({\bf A}\,{\bf B}+{\bf B}\,{\bf A})\,,
\endeq
where we indicate the scalar part by the subscript ``$S$''.
Accordingly we define the exterior product by the commutator
\begeq
{\bf A}\wedge{\bf B}\equiv{{\bf A}\,{\bf B}-{\bf B}\,{\bf A}\over 2}\,.
\endeq
Now that we defined the products, we should come back to the unit vectors.
The only ``unit vector'' that we explicitely defined so far is the symplectic 
unit matrix $\y_0$. If it represents anything at all then it must be 
``the direction'' of change, the direction of evolution in time as it was
derived in this context and is the only ``dimension'' found so far. 
As we have already shown, all other unit vectors $\y_k$ must be symmetric, 
if they are symplices.
And vice versa: If $\y_k$ is symmetric and anti-commutes with $\y_0$, then it is 
a symplex. As only symplices represent observables and are generators of 
symplectic transformations, we can have only a single ``time'' direction 
$\y_0$ and a yet unknown number of {\it symmetric} unit vectors~\footnote{
Thus we found a simple answer to the question, why only a single time
direction is possible, a question also debated in Ref.~\cite{BN2000}}.
However, for $n>1$, there might be different equivalent choices of $\y_0$. 
Whatever the specific form of $\y_0$ is, we will show that in combination 
with some general requirements like completeness, normalizability and 
observability it determines the structure of the complete algebra. 
Though we don't yet know how many symmetric and pairwise
anti-commuting unit vectors $\y_k$ exist - we have to interpret them as
unit vectors in ``spatial directions''~\footnote{The meaning of what a spatial
direction is, especially in contrast to the direction of time $\y_0$,
has to be derived from the form of the emerging equations, of course.
As meaning follows form, we do not define space-time, but we identify
structures that fit to the known concept of space-time.}.
Of course unit vectors must have unit length, so that we have to demand that
\begary{rcl}
\Vert\y_k\Vert^2={1\over 2\,n}\mathrm{Tr}(\y_k^2)=\pm 1\,.
\endary
Note that (since our norm is not positive definite), we explicitely allow
for unit vectors with negative ``length'' as we find it for $\y_0$. 
Note furthermore that all skew-symmetric unit vectors square to 
$-{\bf 1}$ while the symmetric ones square to ${\bf 1}$~\cite{qed_paper}.

Indeed systems of $N=p+q$ anti-commuting real matrices are known as 
real representations of Clifford algebras $Cl_{p,q}$. The index $p$ is the
number of unit elements (``vectors'') that square to $+1$ and $q$ is the 
number of unit vectors that square to $-1$. 
Clifford algebras are not necessarily connected to Hamiltonian motion, rather
they can be regarded as purely mathematical ``objects''. They can be defined without
reference to matrices whatsoever. Hence in mathematics, sets of matrices are merely
``representations'' of Clifford algebras. But our game is about physics and
due to the proven one-dimensionality of time we concentrate on Clifford
algebras $Cl_{N-1,1}$ which link CHOs in the described way with the generators
of a Minkowski type metric.
Further below it will turn out that the representation by matrices is - within the
game - indeed helpful, since it leads to an overlap of certain symmetry structures.
The unit elements (or unit ``vectors'') of a Clifford algebra, ${\bf e}_k$, are 
called the {\it generators} of the Clifford algebra. They pairwise anticommute
and they square to $\pm{\bf 1}$~\footnote{
The role as {\it generator} of the Clifford algebra should not be confused with the 
role as generators of symplectic transformations (i.e. symplices). Though we
are especially interested in Clifford algebras in which all generators are
symplices, not all symplices are generators of the Clifford algebra. 
Bi-vectors for instance are symplices, but not generators of the Clifford algebra.}.
Since the inverse of the unit elements ${\bf e}_k$ of a Clifford algebra must be unique, 
the products of different unit vectors form new elements and all possible
products including the unit matrix form a group. There are $\left({N\atop k}\right)$ 
possible combinations (products without repetition) of $k$ elements from a set of $N$ 
generators. We therefore find $\left({N\atop 2}\right)$ {\it bi-vectors}, which are 
products of 2 generators, $\left({N\atop 3}\right)$ {\it trivectors}) and so on.
The product of all $N$ basic matrices is called {\it pseudoscalar}.
The total number of all k-vectors then is~\footnote{We identify $k=0$ with 
the unit matrix ${\bf 1}$.}:
\begeq
\sum\limits_{k=0}^N\,\left({N\atop k}\right)=2^N\,.
\endeq
If we desire to construct a {\it complete} system, then the number of
variables of the Clifford algebra has to match the number of variables of 
the used matrix system:
\begeq
2^N=(2n)^2\,.
\label{eq_complete}
\endeq
Note that the root of this equation gives an even integer $2^{N/2}=2\,n$ so
that $N$ must be even. Hence all Hamiltonian Clifford algebras have an even
number of dimensions. Of course not all elements of the Clifford algebra are
symplices. The unit matrix (for instance) is a cosymplex. Consider the 
Clifford algebra $Cl_{1,1}$ with $N=2$, which has two generators,
say $\y_0$ with $\y_0^2=-{\bf 1}$ and $\y_1$ with $\y_1^2={\bf 1}$. 
Since these two anticommute (by definition of the Clifford algebra), so that
we find (besides the unit matrix) a fourth matrix formed by the product
$\y_0\,\y_1$:
\begary{rcl}
\y_0\,\y_1&=&-\y_1\,\y_0\\
(\y_0\,\y_1)^2&=&\y_0\,\y_1\,\y_0\,\y_1\\
              &=&-\y_0\,\y_0\,\y_1\,\y_1={\bf 1}\,.
\endary
The completeness of the Clifford algebras as we use them here implies
that any $2\,n\times 2\,n$-matrix ${\bf M}$ with $(2n)^2=2^N$ can be 
written as a linear combination of all elements of the Clifford algebra:
\begeq
{\bf M}=\sum\limits_{k=0}^{4n^2-1}\,m_k\,\y_k\,.
\label{eq_completeness}
\endeq
The coefficients can be computed from the scalar product of the unit
vectors with the matrix ${\bf M}$:
\begeq
m_k=(\y_k\cdot{\bf M})_S={s_k\over 4n}\,\mathrm{Tr}(\y_k\,{\bf M}+{\bf M}\,\y_k)\,.
\label{eq_coeffs}
\endeq
Recall that skew-symmetric $\y_k$ have a negative length and therefore
we included a factor $s_k$ which represents the ``signature'' of $\y_k$, 
in order to get the correct sign of the coefficients $m_k$.

Can we derive more properties of the constructable space-times? One
restriction results from representation theory: A theorem from the
theory of Clifford algebras states that $Cl_{p,q}$ has a representation 
by real matrices if (and only if)~\cite{Lounesto}
\begeq
p-q=0\textrm{ or }2\textrm{ mod }8\,.
\endeq
The additional requirement that all generators must be symplices so that
$p=N-1$ and $q=1$ then restricts $N$ to
\begeq
N-2=0\textrm{ or }2\textrm{ mod }8\,.
\label{eq_dim}
\endeq
Hence the only matrix systems that have the required symmetry properties 
within our modeling game are those that represent Clifford algebras with
the dimensions $1+1$, $3+1$, $9+1$, $11+1$, $17+1$, 
$19+1$, $25+1$, $27+1$ and so on. These correspond to matrix representations
of size $2\times 2$, $4\times 4$, $32\times 32$, $64\times 64$, $512\times 512$
and so on. The first of them is called {\it Pauli algebra}, the second one is the
{\it Dirac algebra}. Do these two have special properties that the
higher-dimensional algebras do not have? Yes, indeed. 

Firstly, since dynamics 
is based on canonical pairs, the real Pauli algebra describes the motion of a 
single DOF and the Dirac algebra decribes the {\it simplest system 
with interaction} between two DOF. This suggests the interpretation that
within our game, objects (Dirac-particles) are not located ``within
space-time'', since we did not define space {\it at all} up to this point, but
that space-time can be modeled as an emergent phenomenon. 

Secondly, if we equate the number of fundamental variables ($2\,n$) of the 
oscillator phase space with the dimension of the Clifford space $N$, then 
Eqn.~\ref{eq_complete} leads to
\begeq
2^N=N^2\,,
\label{eq_xcomplete}
\endeq
which allows for $N=2$ and $N=4$ only. But why should it be meaningful to assume
$N=2\,n$? The reason is quite simple: If $2\,n>N$ as for all
higher-dimensional state vectors, there are not enough generators of the
algebra as there are variables. This discrepancy increases with $n$. Hence
the described objects can not be pure {\it vectors} anymore, but must contain
tensor-type components ($k$-vectors)~\footnote{For a deeper discussion of
the dimensionality of space-time, see Ref.~\cite{qed_paper} and references
therein.}.

But before we describe a formal way to interprete Eqn.~\ref{eq_xcomplete}, 
let us first investigate the physical and geometrical implications of the 
game as described so far.

\subsection{Matrix Exponentials}

We said that the unit vectors $\y_0$ and $\y_k$ are symplices and therefore 
generators of symplectic transformations. All symplectic matrices are 
matrix exponentials of symplices~\cite{MHO}. The computation
of matrix exponentials is in the general case non-trivial. However, in the 
special case of matrices that square to $\pm{\bf 1}$~\footnote{E.g. along the
``axis'' $\y_k$ of the coordinate system.}, the exponentials are 
readily evaluated:
\begary{rcl}
\exp{(\y_a\,\tau)}&=&\sum\limits_{k=0}^\infty\,{(\y_a\,\tau)^k\over k!}\\
\exp{(\y_a\,\tau)}&=&\sum\limits_{k=0}^\infty\,s^k\,{\tau^{2k}\over (2\,k)!}
+\y_a\,\sum\limits_{k=0}^\infty\,s^k\,{\tau^{2k+1}\over (2\,k+1)!}\,,
\endary
where $s=\pm 1$ is the sign of the matrix square of $\y_a$. For $s=-1$ ($\y_a^2=-{\bf 1}$), 
it follows that
\begeq
{\bf R}_a(\tau)=\exp{(\y_a\,\tau)}=\cos{(\tau)}+\y_a\,\sin{(\tau)}\,,
\label{eq_rot}
\endeq
and for $s=1$ ($\y_a^2={\bf 1}$):
\begeq
{\bf B}_a(\tau)=\exp{(\y_a\,\tau)}=\cosh{(\tau)}+\y_a\,\sinh{(\tau)}\,,
\label{eq_boost}
\endeq
We can indentify skew-symmetric generators with rotations and (as we
will show in more detail below) symmetric generators with boosts.

The (hyperbolic) sine/cosine structure of symplectic matrices are not limited
to the generators but are a general property of the matrix exponentials of
the symplex ${\bf F}$~\footnote{These properties are the main motivation to
choose the nomenclature of ``symplex'' and ``cosymplex''.}:
\begeq
{\bf M}(t)=\exp{({\bf F}\,t)}={\bf C}+{\bf S}\,,
\endeq
where the (co-) symplex ${\bf S}$ (${\bf C}$) is given by:
\begary{rcl}
{\bf S}&=&\sinh{({\bf F}\,t)}\\
{\bf C}&=&\cosh{({\bf F}\,t)}\,,
\endary
since (the linear combination of) all odd powers of a symplex is again a
symplex and the sum of all even powers is a cosymplex.
The inverse transfer matrix ${\bf M}^{-1}(t)$ is given by:
\begeq
{\bf M}^{-1}(t)={\bf M}(-t)={\bf C}-{\bf S}\,.
\endeq
The physical meaning of the matrix exponential results from
Eqn.~\ref{eq_eqom_cho}, which states that (for constant symplices ${\bf F}$)
the solutions are given by the matrix exponential of ${\bf F}$:
\begeq
\psi(t)={\bf M}(t)\,\psi(0)\,.
\label{eq_mtxexp}
\endeq
A symplectic transformation can be regarded as the result of a {\it possible}
evolution in time.
There is no prove that non-symplectic processes are forbidden by nature, 
but that {\it only} symplectic transformations are {\it structure preserving}.
Non-symplectic transformations are then {\it structure defining}. 
Both play a fundamental role in the physics of our model reality, {\it
  because} fundamental particles are - according to our model - represented
by dynamical structures. Therefore symplectic transformations describe
those processes and interactions, in which structure is preserved, i.e.
in which the type of the particle is not changed. The fundamental variables 
are just ``carriers'' of the dynamical structures. 
Non-symplectic transformations can be used to transform the structure. This
could also be described by a rotation of the direction of time. Another 
interpretation is that of a gauge-transformation~\cite{newlook}.

\section{The Significance of (De-)Coupling}
\label{sec_decoupling}

In physics it is a standard technique to reduce complexity of problems
by a suitable change of variables. In case of linear systems, the change
of variables is a linear canonical transformation. The goal of such
transformations is usually to substitute the solution of a complicated
problem by the solution of multiple simpler systems. This technique
is known under various names, one of these names is {\it decoupling}, 
but it is also known as {\it principal component analysis} or (as we will
later show) transformation into the ``rest frame''. In other
branches of science one might refer to it as {\it pattern recognition}. 

In the following we investigate, how to compute (or recognize) the eigenvectors
and eigenvalues of a general $2n\times 2n$-dimensional symplex. Certainly it 
would be preferable to find a ``physical method'', i.e. a method that matches
to the concepts that we introcuded so far and that has inherently physical 
significance. Or at least significance and explantory power with respect to 
our modeling game. Let us start from the simplest systems, i.e. with the Pauli
and Dirac algebras which correspond to matrices of size $2\times 2$ and
$4\times 4$, respectively.
 
\subsection{The Pauli Algebra}
\label{sec_pauli}

The fundamental significance of the Pauli algebra is based on the even
dimensionality of (classical) phase space. The algebra of $2\times 2$ 
matrices describes the motion of a single (isolated) DOF.
Besides $\eta_0$, the real Pauli algebra includes the following three 
matrices: 
\begary{rcl}
\eta_1&=&\bmtx{cc}0&1\\1&0\\\emtx\\
\eta_2&=&\eta_0\,\eta_1=\bmtx{cc}1&0\\0&-1\emtx\\
\eta_3&=&{\bf 1}=\bmtx{cc}1&0\\0&1\emtx\\
\endary
All except the unit matrix $\eta_3$ are symplices. If $\eta_0$ and $\eta_1$
are chosen to represent the generators of the corresponding Clifford
algebra $Cl_{1,1}$, then $\eta_2$ is the only possible bi-vector.
A general symplex has the form:
\begary{rcl}
{\bf F}&=&a\,\eta_0+b\,\eta_1+c\,\eta_2\\
&=&\bmtx{cc}
c & a+b\\
-a+b&-c\\
\emtx\,.
\endary
The characteristic equation is given by $\mathrm{Det}({\bf F}-\lambda\,{\bf 1})=0$
\begary{rcl}
0&=&(c-\lambda)(-c-\lambda)-(a+b)(-a+b)\\
\lambda&=&\pm\sqrt{c^2+b^2-a^2}\\
\endary
The eigenvalues $\lambda_{\pm}$ are both either real for $a^2<c^2+b^2$ or both
imaginary $a^2>c^2+b^2$ (or both zero). Systems in stable oscillation have purely
imaginary eigenvalues. This case is most interesting for our modeling game.

{\it Decoupling} is usually understood in the more general sense to treat the
interplay of several (at least two) DOF, but here we ask, whether all possible 
oscillating systems of $n=1$ are isomorphic to normal form oscillators.
Since there are $3$ parameters in ${\bf F}$ and only one COM, namely the frequency 
$\w$, we need at least two parameters in the transformation matrix. 
Let us see, if we can choose these two transformations along the axis of 
the Clifford algebra. In this case we apply subsequentially two symplectic 
transformations along the axis $\eta_0$ and $\eta_2$. Applying the symplectic 
transformation matrix $\exp{(\eta_0\,\tau/2)}$ we obtain:
\begary{rcl}
{\bf F}_1&=&\exp{(\eta_0\,\tau/2)}\,{\bf F}\,\exp{(-\eta_0\,\tau/2)}\\
&=&a'\,\eta_0+b'\,\eta_1+c'\,\eta_2\\
\endary~\footnote{The ``half-angle'' argument is for convenience.}
The transformed coefficients $a'$, $b'$ and $c'$ are given by
\begary{rcl}
a'       &=&a\\
b'       &=&b\,\cos{\tau}-c\,\sin{\tau}\\
c'       &=&c\,\cos{\tau}+b\,\sin{\tau}\\
\endary
so that - depending on the ``duration of the pulse'', we can chose to 
transform into a coordinate system in which either $b'=0$ or $c'=0$. If we
choose $t=\arctan{(-c/b)}$, then $c'=0$, so that
\begary{rcl}
{\bf F}'&=&a\,\eta_0+\sqrt{b^2+c^2}\,\eta_1=a'\,\eta_0+b'\,\eta_1\\
\endary
If we chose the next generator to be $\eta_2$, then:
\begary{rcl}
a''       &=&a'\,\cosh{\tau}-b'\,\sinh{\tau}\\
b''       &=&b'\,\cosh{\tau}-a'\,\sinh{\tau}\\
\endary
In this case we have to dinstinguish between the case, where $a'>b'$
and $a'<b'$. The former is the oscillatory system and in this case
the transformation with $\tau=\mathrm{artanh}{(b'/a')}$ leads to the normal form of a 1-dim. oscillator:
\begary{rcl}
a''       &=&\sqrt{a^2-b^2-c^2}\\
b''       &=&0\\
c''       &=&0\\
\endary
and the matrix ${\bf F}''$ has the form
\begeq
{\bf F}''=\sqrt{a^2-b^2-c^2}\,\eta_0\,.
\endeq
If the eigenvalues are imaginary, then $\lambda=\pm\,i\,\omega$ and hence
\begeq
{\bf F}''=\omega\,\eta_0\,,
\label{eq_1dcho}
\endeq
so that the solution is - for constant frequency - given by the 
matrix exponential:
\begary{rcl}
\psi(t)&=&\exp{(\omega\,\eta_0\,t)}\,\psi(0)\\
       &=&\left({\bf 1}\,\cos{(\omega\,t)}+\eta_0\,\sin{(\omega\,t)}\right)\,\psi(0)\,.
\label{eq_1dsolution}
\endary
This shows that in the context of stable oscillator algebras - the real 
Pauli algebra can be reduced to the complex number system: This becomes evident, 
if we consider possible 
representations of the complex numbers. Clearly we need two basic elements - 
the unit matrix and $\eta_0$, i.e. a matrix that commutes with the
unit matrix and squares to $-{\bf 1}$. If we write ``$i$'' instead of 
$\eta_0$, then it is easily verified that~\footnote{
See also Refs.~\cite{Lounesto,Dyson} and Eqn.~\ref{eq_composite} in combination
with Ref.~\cite{Ralston1989}.}:
\begary{rcl}
z&=&x+i\,y={\bf Z}=\bmtx{cc}x&y\\-y&x\emtx\\
\bar z&=&x-i\,y={\bf Z}^T=x\,{\bf 1}+\eta_0^T\,y\\
\exp{(i\,\phi)}&=&\cos{(\phi)}+i\,\sin{(\phi)}\\
\Vert z\Vert^2&=&{\bf Z}\,{\bf Z}^T=z\,\bar z=x^2+y^2\\
\label{eq_complex}
\endary
The theory of holomorphic functions is based on series expansions and
can be equally well formulated with matrices. Viewed from our perspective
the complex numbers are a special case of the real Pauli algebra - 
since we have shown above that any one-dimensional oscillator 
can be canonically transformed into a system of the form of Eqn.~\ref{eq_1dcho}.
Nevertheless we emphasize that the complex numbers interpreted this way 
can only represent the {\it normal form} of an oscillator. The normal 
form excludes a different scaling of coordinates and momenta as used 
in classical mechanics, i.e. it avoids intrinsically the appearance 
of different ``spring constants'' and masses~\footnote{ 
There have been several attempts to explain the appearance of the complex
numbers in quantum mechanics~\cite{StB60,Strocchi,Hest75,Baylis92,
MEH2012,Gibb2012,Schiff2012}. 
A general discussion of the use of complex numbers in physics
is beyond the scope of this essay, therefore we add just a remark. 
Gary W. Gibbons wrote that ``In particular there can be no evolution 
if $\psi$ is real''~\cite{Gibb2012}.
We agree with Gibbons that the unit imaginary can be related to evolution
in time as it implies oscillation, but we do not agree with his conclusion.
Physics was able to describe evolution in time without imaginaries 
before quantum mechanics and it still is. The unconscious use of the unit
imaginary did not prevent quantum mechanics from being experimetally
successful. But it prevents physicists from understanding its structure.}.

\subsection{The Dirac Algebra}
\label{sec_dirac}

In this subsection we consider the oscillator algebra for two coupled DOF, 
the algebra of $4\times 4$ matrices. In contrast to the real Pauli
algebra, where the parameters $a$, $b$ and $c$ did not suggest a specific 
physical meaning, the structure of the Dirac algebra bears geometrical significance
as has been pointed out by David Hestenes and others~\cite{STA1,STA2,GLD}. 
The (real) Dirac algebra is the simplest real algebra that enables for a 
description of two DOF and the interaction between them. 
Furthermore the eigenfrequencies of a general symplex ${\bf F}$
may be complex, while the spectrum of the Pauli matrices does not include 
complex numbers off the real and imaginary axis. 
The spectrum of general $2n\times 2n$-symplices has a certain structure -
since the coefficients of the characteristic polynomial are real: 
If $\lambda$ is an eigenvalue of ${\bf F}$, then its complex conjugate
$\bar\lambda$ as well as $\lambda$ and $-\bar\lambda$ are also eigenvalues.
As we will show, this is the spectrum of the Dirac algebra and therefore
any $2n\times 2n$-system can at least in principle be block-diagonalized 
using $4\times 4$-blocks. The Pauli algebra is therefore not sufficient to 
cover this general case.

The structure of Clifford algebras follows Pascal's triangle. The Pauli
algebra has the structure $1-2-1$ (scalar-vector-bivector), the Dirac 
algebra has the structure $1-4-6-4-1$, standing for unit element (scalar), 
vectors, bi-vectors, tri-vectors and pseudoscalar. The vector elements 
are by convention indexed with $\y_\mu$ with $\mu=0\dots 3$, i.e. the 
generators of the algebra~\footnote{
According to Pauli's fundamental theorem of the Dirac algebra, 
all possible choices of the Dirac matrices are, as long as the ``metric
tensor'' $g_{\mu\nu}$ remains unchanged, physically equivalent~\cite{Pauli}.
}:
{\small
\begeq
\begin{array}{rclp{4mm}rcl}
\y_0&=&\bmtx{cccc}
   0 &   1  &  0 &   0\\
  -1 &   0  &  0 &   0\\
   0 &   0  &  0 &   1\\
   0 &   0  & -1 &   0\\
\emtx&&
 \y_1&=&\bmtx{cccc}
   0 &  -1  &  0 &   0\\
  -1 &   0  &  0 &   0\\
   0 &   0  &  0 &   1\\
   0 &   0  &  1 &   0\\
\emtx\\
 \y_2&=&\bmtx{cccc}
   0 &   0  &  0 &   1\\
   0 &   0  &  1 &   0\\
   0 &   1  &  0 &   0\\
   1 &   0  &  0 &   0\\
\emtx&&
 \y_3&=&\bmtx{cccc}
  -1 &   0  &  0 &   0\\
   0 &   1  &  0 &   0\\
   0 &   0  & -1 &   0\\
   0 &   0  &  0 &   1\\
\emtx\\
\end{array}\,.
\endeq}
We define the following numbering scheme for the remaining matrices~\footnote{
The specific choice of the matrices is not unique. A survey of the different
systems can be found in Ref.(~\cite{rdm_paper}).}:
{\small
\begeq
\begin{array}{rclp{4mm}rcl}
\y_{14}&=&\y_0\,\y_1\,\y_2\,\y_3;&&\y_{15}&=&{\bf 1}\\
\y_4&=&\y_0\,\y_1;&&\y_7&=&\y_{14}\,\y_0\,\y_1=\y_2\,\y_3\\
\y_5&=&\y_0\,\y_2;&&\y_8&=&\y_{14}\,\y_0\,\y_2=\y_3\,\y_1\\
\y_6&=&\y_0\,\y_3;&&\y_9&=&\y_{14}\,\y_0\,\y_3=\y_1\,\y_2\\
\y_{10}&=&\y_{14}\,\y_0&=&\y_1\,\y_2\,\y_3&&\\
\y_{11}&=&\y_{14}\,\y_1&=&\y_0\,\y_2\,\y_3&&\\
\y_{12}&=&\y_{14}\,\y_2&=&\y_0\,\y_3\,\y_1&&\\
\y_{13}&=&\y_{14}\,\y_3&=&\y_0\,\y_1\,\y_2&&\\
\end{array}\,.
\label{eq_diracalgebra}
\endeq}
According to Eq.~\ref{eq_ns} we expect $10$ symplices and since
the $4$ vectors and $6$ bi-vectors are symplices, all other elements
are cosymplices. With this ordering, the general $4\times 4$-symplex 
${\bf F}$ can be written as (instead of Eq.~\ref{eq_completeness}):
\begeq
{\bf F}=\sum\limits_{k=0}^9\,f_k\,\y_k\,.
\label{eq_f_coeffs}
\endeq
In Ref.~\cite{rdm_paper} we presented a detailed survey of the Dirac
algebra with respect to symplectic Hamiltonian motion. The essence of
this survey is the insight that the real Dirac algebra describes 
Hamiltonian motion of an ensembles of two-dimensional
oscillators, but as well the motion of a ``point particle'' in
3-dimensional space, i.e. that Eqn.~\ref{eq_envelope} is, when
expressed by the real Dirac algebra, {\it isomorphic to the Lorentz
force equation} as we are going to show in Sec.~\ref{sec_LorentzForce}.
Or, in other words, the Dirac algebra allows to 
model a point particle and its interaction with an electromagnetic
field in terms of the classical statistical ensemble of abstract 
oscillators.

\section{Electromechanical Equivalence (EMEQ)}
\label{sec_emeq}

The number and type of symplices within the Dirac
algebra~(\ref{eq_diracalgebra}) suggests to use the following 
vector notation for the coefficients~\cite{rdm_paper,geo_paper} of the
observables:
\begary{rcl}
{\cal E}&\equiv&f_0\\
\vec P&\equiv&(f_1,f_2,f_3)^T\\
\vec E&\equiv&(f_4,f_5,f_6)^T\\
\vec B&\equiv&(f_7,f_8,f_9)^T\,,
\label{eq_emeq}
\endary
where the ``clustering'' of the coefficients into 3-dimensional vectors 
will be explained in the following. The first four elements ${\cal E}$ and
$\vec P$ are the coefficients of the generators of the Clifford algebra and
the remaining symplices are $3$ symmetric bi-vectors $\vec E$ and skew-symmetric
bi-vectors $\vec B$. As explained above, the matrix exponentials of pure Clifford
elements are readily evaluated (Eq.~\ref{eq_rot} and Eq.~\ref{eq_boost}).
The effect of a symplectic similarity transformation on a symplex 
\begary{rcl}
\tilde\psi&=&{\bf R}(\tau/2)\,\psi\\
{\bf\tilde F}&=&{\bf R}(\tau/2)\,{\bf F}\,{\bf R}^{-1}(\tau/2)\\
             &=&{\bf R}(\tau/2)\,{\bf F}\,{\bf R}(-\tau/2)\\
\endary
can then be evaluated component-wise as in the following case of a rotation (using Eq.~\ref{eq_f_coeffs}):
\begary{rcl}
{\bf\tilde F}&=&\sum\limits_{k=0}^9\,f_k\,{\bf R}_a\,\y_k\,{\bf R}_a^{-1}\\
{\bf R}_a\,\y_k\,{\bf R}_a^{-1}&=&\left(\cos(\tau/2)+\y_a\,\sin(\tau/2)\right)\,\y_k\,\\
                               &\times&\left(\cos(\tau/2)-\y_a\,\sin(\tau/2)\right)\\
                               &=&\y_k\,\cos^2(\tau/2)-\y_a\,\y_k\,\y_a\,\sin^2(\tau/2)\\
                               &+&(\y_a\,\y_k-\y_k\,\y_a)\,\cos(\tau/2)\,\sin(\tau/2))\\
\endary
Since all Clifford elements either commute or anti-commute with each other,
we have two possible solutions. The first ($\y_k$ and $\y_a$ commute) yields with $\y_a^2=-{\bf 1}$:
\begeq
{\bf R}_a\,\y_k\,{\bf R}_a^{-1}=\y_k\,\cos^2(\tau/2)-\y_a^2\,\y_k\,\sin^2(\tau/2)=\y_k\,,
\endeq
but if ($\y_k$ and $\y_a$ anti-commute) we obtain a rotation:
\begary{rcl}
{\bf R}_a\,\y_k\,{\bf R}_a^{-1}&=&\y_k\,(\cos^2(\tau/2)-\sin^2(\tau/2))\\
                               &+&\y_a\,\y_k\,2\,\cos(\tau/2)\,\sin(\tau/2))\\
                               &=&\y_k\,\cos(\tau)+\y_a\,\y_k\,\sin(\tau)\,.
\endary
For $a=9$ ($\y_a=\y_1\,\y_2$) for instance we find:
\begary{rcl}
\tilde\y_1&=&\y_1\,\cos(\tau)+\y_1\,\y_2\,\y_1\,\sin(\tau)=\y_1\,\cos(\tau)-\y_2\,\sin(\tau)\\
\tilde\y_2&=&\y_2\,\cos(\tau)+\y_1\,\y_2\,\y_2\,\sin(\tau)=\y_2\,\cos(\tau)+\y_1\,\sin(\tau)\\
\tilde\y_3&=&\y_3\,,
\endary
which is formally equivalent to a rotation of $\vec P$ about the ``z-axis''.
If the generator $\y_a$ of the transformation is symmetric, we obtain:
\begary{rcl}
{\bf R}_a\,\y_k\,{\bf R}_a^{-1}&=&(\cosh(\tau/2)+\y_a\,\sinh(\tau/2))\,\y_k\,\\
                               &\times&(\cosh(\tau/2)-\y_a\,\sinh(\tau/2))\\
                               &=&\y_k\,\cosh^2(\tau/2)-\y_a\,\y_k\,\y_a\,\sinh^2(\tau/2)\\
                               &+&(\y_a\,\y_k-\y_k\,\y_a)\,\cosh(\tau/2)\,\sinh(\tau/2))\,,
\endary
so that (if $ \y_a$ and $\y_k$ commute):
\begary{rcl}
\tilde\y_k&=&\y_k\,\cosh^2(\tau/2)-\y_a^2\,\y_k\,\sinh^2(\tau/2)\\
\tilde\y_k&=&\y_k\,(\cosh^2(\tau/2)-\sinh^2(\tau/2))=\y_k\\
\endary
and if $ \y_a$ and $\y_k$ anticommute:
\begary{rcl}
\tilde\y_k&=&\y_k\,(\cosh^2(\tau/2)+\sinh^2(\tau/2))\\&+&2\,\y_a\,\y_k\,\cosh(\tau/2)\,\sinh(\tau/2))\\
          &=&\y_k\,\cosh(\tau)+\y_a\,\y_k\,\sinh(\tau)\,,
\endary
which is formally equivalent to a boost, as it allows for the following parametrization
of ``rapidity'' $\tau$:
\begary{rcl}
\tanh{(\tau)}&=&\beta\\
\sinh{(\tau)}&=&\beta\,\gamma\\
\cosh{(\tau)}&=&\gamma\\
\gamma&=&{1\over\sqrt{1-\beta^2}}\,.
\endary
A complete survey of these transformations and the (anti-) commutator tables can be
found in Ref.~\cite{rdm_paper}. However this formalism corresponds exactly to the 
relativistic invariance of a Dirac spinor in QED as described for instance in 
Ref.~\cite{Schmueser}.
The ``spatial'' rotations are generated by the bi-vectors associated with $\vec B$ 
and Lorentz boosts by the components associated with $\vec E$.
The remaining $4$ generators of symplectic transformations correspond to ${\cal E}$
and $\vec P$. They where named {\it phase-rotation} (generated by $\y_0$) and 
{\it phase-boosts} (generated by $\vec\y=(\y_1,\y_2,\y_3)$) and have been used
for instance for symplectic decoupling as described in Ref.~\cite{geo_paper}.

It is nearby (and already suggested by our notation) to consider the possibility
that the EMEQ (Eq.~\ref{eq_emeq}) allows to model a relativistic particle as 
represented by energy ${\cal E}$ and momentum ${\bf P}$ either in an external
electromagnetic field given by $\vec E$ and $\vec B$ or - alternatively - in an 
accelerating and/or rotating reference frame, where the elements $\vec E$ and $\vec B$ 
correspond to the axis of acceleration and rotation, respectively. We assumed
at beginning, that all components of the state vector $\psi$ are equivalent
in meaning and unit. Though we found that the state vector is formally composed
of canonical pairs, the units are unchanged and identical for all elements
of $\psi$. From Eq.~\ref{eq_eqom_cho} we take, that the simplex ${\bf F}$ (and
also ${\cal A}$) have the unit of a frequency.
If the Hamiltonian ${\cal H}$ is supposed to represent energy, then the 
components of $\psi$ have the unit of the square root of action.

If the coefficients are supposed to represent electromagnetic field, then
we need to express these fields in units of frequency. This can be done, 
but it requires to involve natural conversion factors like $\hbar$, 
charge $e$, velocity $c$ and a mass, for instance the electron mass $m_e$.
The magnetic field (for instance) is related to a ``cyclotron frequency'' 
$\omega_c$ by $\omega_c\propto {e\over m_e}\,B$.

However, according to the rules of the game, the distinction between particle
properties and ``external'' fields requires a reason, an explanation. 
Especially as it is physically meaningless for macroscopic coupled oscillators. 
In Refs.~\cite{rdm_paper,geo_paper}, we used this nomenclature in a merely 
{\it formal} way, namely to find a descriptive scheme to order
the symplectic generators, so to speak an {\it equivalent circuit} to 
describe the general possible coupling terms for two-dimensional coupled 
linear optics as required for the description of charged particles beams.
 
Here we play the reversed modeling game: Instead of using the EMEQ as an
equivalent circuit to describe ensembles of oscillators, we now
use ensembles of oscillators as an equivalent circuit to describe point 
particles. The motivation for Eqn.~\ref{eq_emeq} is nevertheless similar, 
i.e. it follows the formal structure of the Dirac Clifford algebra. The 
grouping of the coefficients comes along with the number of vector- and 
bi-vector-elements, $4$ and $6$, respectively. The second criterium is to 
distinguish between generators of rotations and boost, i.e. between 
symmetric and skew-symmetric symplices, which separates energy from 
momentum and electric from magnetic elements. 
Third of all, we note that even~\footnote{{\it Even} $k$-vectors are those
with even $k=2\,m$, where $m$ is a natural number.} elements (scalar, 
bi-vectors, $4$-vectors etc.) of even-dimensional Clifford algebras form 
a sub-algebra. This means that we can generate the complete Clifford algebra 
from the vector-elements by matrix multiplication (this is why we call them
generators), but we can not generate vectors from bi-vectors 
by multiplication. And therefore the vectors are the particles (which are 
understood as the sources of fields) and the bi-vectors are the fields, 
which are generated by the objects and influence their motion. 
The full Dirac symplex-algebra includes the description of a particle (vector) 
in a field (bi-vector).
But why would the field be {\it external}? Simply, because it is impossible
to generate bi-vectors from a single vector-type object, since any single
vector-type object written as ${\cal E}\,\y_0+\vec P\cdot\vec\y$ squares
to a scalar. Dirac aimed for this result when he invented the Dirac matrices.
Therefore, the fields must be the result of interaction with other particles
and hence we call them ``external''. This is in some way a ``first-order''
approach, since there might be higher order processes that we did not
consider yet. But in the linear approach (i.e. for second-order Hamiltonians),
this distinction is reasonable and hence a legitimate move in the game.

Besides the Hamiltonian structure (symplices vs. co-symplices) and
the Clifford algebraic structure (distinguishing vectors, bi-vectors, 
tri-vectors etc) there is a third essential symmetry, which is connected 
to the real matrix representation of the Dirac algebra and to the fact
that it describes the general Hamiltonian motion of coupled oscillators:
To distinguish the even from the odd elements with respect to the
block-diagonal matrix structure. We used this property in Ref.~\cite{
geo_paper} to develop a general geometrical decoupling algorithm (see
also Sec.~\ref{sec_geom}).

Now it may appear that we are cheating somehow, as relativity is usually
``derived'' from the constancy of the speed of light, while in our modeling
game, we did neither introduce spatial notions nor light at all. Instead
we directly arrive at notions of quantum electrodynamics (QED). 
How can this be? The definition of ``velocity'' within wave mechanics
usually involves the dispersion relation of waves, i.e. the velocity of a 
wave packet is given by the group velocity $\vec v_{gr}$ defined by
\begeq
\vec v_{gr}\equiv \vec\nabla_{\vec k}\,\omega(\vec k)\,,
\label{eq_vgr}
\endeq
and the so-called phase velocity $v_{ph}$ defined by
\begeq
v_{ph}={\omega\over  k}\,.
\label{eq_vph}
\endeq
It is then typically mentioned that the product of these two velocities
is a constant $v_{gr}\,v_{ph}=c^2$.
By the use of the EMEQ and Eq.~\ref{eq_laxconst}, the eigenvalues of ${\bf F}$ 
can be written as:
\begary{rcl}
K_1&=&-\mathrm{Tr}({\bf F}^2)/4\\
K_2&=&\mathrm{Tr}({\bf F}^4)/16-K_1^2/4\\
\omega_1 &=&\sqrt{K_1+2\,\sqrt{K_2}}\\
\omega_2 &=&\sqrt{K_1-2\,\sqrt{K_2}}\\
\w_1^2\,\w_2^2&=&K_1^2-4\,K_2=\mathrm{Det}({\bf F})\\
K_1&=&{\cal E}^2+\vec B^2-\vec E^2-\vec P^2\\
K_2&=&({\cal E}\,\vec B+\vec E\times\vec P)^2-(\vec E\cdot\vec B)^2-(\vec P\cdot\vec B)^2\\
\label{eq_eigenfreq}
\endary
Since symplectic transformations are similarity transformations, they do not
alter the eigenvalues of the matrix ${\bf F}$ and since all possible evolutions
in time (which can be described by the Hamiltonian) are symplectic transformations, 
the eigenvalues (of closed systems) are conserved. If we consider a ``free particle'',
the we obtain from Eq.~\ref{eq_eigenfreq}:
\begeq
\omega_{1,2}=\pm\sqrt{{\cal E}^2-\vec P^2}\,.
\label{eq_energy}
\endeq
As we mentioned before both, energy and momentum, have (within this game) the
unit of frequencies. If we take into account that $\omega_{1,2}\equiv m$ is fixed,
then the dispersion relation for ``the energy'' ${\cal E}=\omega$ is
\begeq
{\cal E}=\omega=\sqrt{m^2+\vec P^2}\,.
\label{eq_dispersion}
\endeq
which is indeed the correct relativistic dispersion.
But how do we make the step from pure oscillations to {\it waves}?~\footnote{
The question if Quantum theory requires Planck's constant $\hbar$, has been
answered negative by John P. Ralston~\cite{Ralston2012}.}.

\subsection{Moments and The Fourier Transform}
\label{sec_density}

In case of ``classical'' probability distribution functions (PDFs) $\phi(x)$ we may use the
Taylor terms of the {\it characteristic function} $\tilde\phi_x(t)=\langle \exp{i\,t\,x}\rangle_x$, 
which is the Fourier transform of $\phi(x)$, at the origin. The $k$-th moment is then given by
\begeq
\langle x^k\rangle=i^k\,\tilde\phi^{(k)}(0)\,.
\endeq
where $\phi^{(k)}$ is the $k$-th derivative of $\tilde\phi_x(t)$.

A similar method would be of interest for our modeling game. Since a (phase space-) density
is positive definite, we can always take the square root of the density instead of the
density itself: $\phi=\sqrt{\rho}$. The square root can also defined to be a complex function,
so that the density is $\rho=\phi\phi^\star=\Vert\phi\Vert^2$ and, if mathematically well-defined (convergent), 
we can also define the Fourier transform of the complex root, i.e.
\begeq
\tilde\phi(\omega,\vec k)=N\,\int\phi(t,\vec x)\,\exp{(i\,\w\,t-i\,\vec k\vec x)}\,dt\,d^3 x\,.
\endeq
and vice versa:
\begeq
\tilde\phi(t,\vec x)=\tilde N\,\int\phi(\w,\vec k)\,\exp{(-i\,\omega\,t+i\,\vec k\vec x)}\,d\w d^3 k\,.
\endeq
In principle, we may {\it define} the density no only by real and imaginary part, but by
an arbitrary number of components. Thus, if we consider a four-component spinor, we may
of course mathematically define its Fourier transform. But in order to see, why this
might be more than a mathematical ``trick'', but {\it physically meaningful}, we need to 
go back to the notions of classical statistical mechanics. Consider that we replace 
the single state vector by an ``ensemble'', where
we leave the question open, if the ensemble should be understood as a single 
phase space trajectory, averaged over time, or as some (presumably large) number of
different trajectories. It is well-known, that the phase space density $\rho(\psi)$
is stationary, if it depends only on constants of motion, for instance if it depends
only on the Hamiltonian itself. With the Hamiltonian of Eq.~\ref{eq_cho_hamiltonian},
the density could for example have the form 
\begeq
\rho({\cal H})\propto\exp{(-\beta\,{\cal H})}=\exp{(-\beta\,\psi\,{\cal A}\,\psi)}\,,
\endeq
which corresponds to a multivariate Gaussian. But more important is the insight, that
the density exclusively depends on the second moments of the phase space variables as
given by the Hamiltonian, i.e. in case of a ``free particle'' it depends on ${\cal E}$
and ${\vec P}$. And therefore we should be able to use energy and momentum as 
frequency $\w$ and wave-vector $\vec k$.

But there are more indications in our modeling game that suggest the use of a Fourier
transform as we will show in the next section.

\subsection{The Geometry of (De-)Coupling}
\label{sec_geom}

In the following we give a (very) brief summary of Ref.~\cite{geo_paper}. As already
mentioned, decoupling is meant - despite the use of the EMEQ - first of
all purely technical-mathematical. Let us delay the question, if the notions 
that we define in the following have any physical relevance. Here we refer
first of all to block-diagonalization, i.e. we treat the symplex ${\bf F}$ 
just as a ``Hamiltonian'' matrix. From the definition of the real Dirac matrices
we obtain ${\bf F}$ in explicit $4\times 4$ matrix form:
{\small\begary{rcl}
{\bf F}&=&\bmtx{cccc}
-E_x&E_z+B_y&E_y-B_z&B_x\\
E_z-B_y&E_x&-B_x&-E_y-B_z\\
E_y+B_z&B_x&E_x&E_z-B_y\\
-B_x&-E_y+B_z&E_z+B_y&-E_x\\
\emtx\\
&+&\bmtx{cccc}
-P_z&{\cal E}-P_x&0&P_y\\
-{\cal E}-P_x&P_z&P_y&0\\
0&P_y&-P_z&{\cal E}+P_x\\
P_y&0&-{\cal E}+P_x&P_z\\
\emtx\,,
\label{eq_edeq1}
\endary}
If we find a (sequence of) symplectic similarity transformations that
would allow to reduce the $4\times 4$-form to a block-diagonal form,
then we would obtain two separate systems of size $2\times 2$ and 
we could continue with the transformations of Sec.~\ref{sec_pauli}.

Inspection of Eqn.~\ref{eq_edeq1} unveils that ${\bf F}$ is block-diagonal, 
if the coefficents $E_y$, $P_y$, $B_x$ and $B_z$ vanish. Obviously 
this implies that $\vec E\cdot\vec B=0$ and $\vec P\cdot\vec B=0$. 
Or vice versa, if we find a symplectic method that transforms into a
system in which $\vec E\cdot\vec B=0$ and $\vec P\cdot\vec B=0$, then 
we only need to apply appropriate rotations to achieve block-diagonal form.
As shown in Ref.~\cite{geo_paper} this can be done in different ways, but
in general it requires the use of the ``phase rotation'' $\y_0$ and 
``phase boosts'' $\vec\y$. Within the conceptional framework of our
game, the application of these transformations equals the use of 
``matter fields''. But furthermore, this shows that block-diagonalization 
has also geometric significance within the Dirac algebra and - with
respect to the Fourier transformation, the requirement $\vec P\cdot\vec B=0$
indicates a divergence free magnetic field, as the replacement
of $\vec P$ by $\vec\nabla$ yields $\vec\nabla\cdot\vec B=0$.
The additional requirement $\vec E\cdot\vec B=0$ also fits well to
our physical picture of e.m. waves. Note furthermore, that there is
no analogous requirement to make $\vec P\cdot\vec E$ equal to zero.
Thus (within this analogy) we {\it can accept} $\vec\nabla\vec E\ne 0$.

But this is not everything to be taken from this method. If we analyze
in more detail, which expressions are {\it required} to vanish and which
may remain, then it appear that $\vec P\cdot \vec B$ is explicitely
given by
\begary{l}
P_x\,B_x\,\y_1\,\y_2\,\y_3+P_y\,B_y\,\y_2\,\y_3\,\y_1+P_z\,B_z\,\y_3\,\y_1\,\y_2\\=(\vec P\cdot \vec B)\,\y_{10}\\
E_x\,B_x\,\y_4\,\y_2\,\y_3+E_y\,B_y\,\y_5\,\y_3\,\y_1+E_z\,B_z\,\y_6\,\y_1\,\y_2\\=(\vec E\cdot \vec B)\,\y_{14}\\
P_x\,E_x\,\y_1\,\y_4\,\y_3+P_y\,E_y\,\y_2\,\y_5\,\y_1+P_z\,E_z\,\y_3\,\y_6\,\y_2\\=-(\vec P\cdot \vec E)\,\y_{0}\,.
\endary
That means that exactly those products have to vanish which yield {\it cosymplices}.
This can be interpreted via the structure preserving properties of symplectic motion. 
Since within our game, the particle {\it type} can only be represented by the structure 
of the dynamics, and since electromagnetic processes do not change the type of a particle, 
then they are quite obviously {\it structure preserving} which then implies
the non-appearance of co-symplices. Or - in other words - electromagnetism is of 
Hamiltonian nature. We will come back to this point in 
Sec.~\ref{sec_maxwell}.

\subsection{The Lorentz Force}
\label{sec_LorentzForce}

In the previous section we constituted the distinction between the 
``mechanical'' elements ${\bf P}={\cal E}\,\y_0+\vec \y\cdot\vec P$ 
of the general matrix ${\bf F}$ and the electrodynamical elements
${\bf F}=\y_0\,\vec\y\cdot E+\y_{14}\,\y_0\,\vec\y\cdot\vec B$. 
Since the matrix ${\bf S}=\Sigma\,\y_0$ is a symplex, let us assume
to be equal to ${\bf P}$ and apply Eqn.~\ref{eq_envelope}. We then find 
(with the appropriate relative scaling between ${\bf P}$ and ${\bf F}$ 
as explained above):
\begeq
{d{\bf P}\over d\tau}=\dot{\bf P}={q\over 2\,m}\left({\bf F}\,{\bf P}-{\bf P}\,{\bf F}\right)\,,
\label{eq_lorentzforce}
\endeq
which yields written with the coefficients of the real Dirac matrices:
\begary{rcl}
{d{\cal E}\over d\tau}&=&{q\over m}\,\vec p\cdot\vec E\\
{d\vec p\over d\tau}&=&{q\over m}\,\left({\cal E}\,\vec E+\vec p\times\vec B\right)\\
\label{eq_lorentzforce1}
\endary
where $\tau$ is the proper time. If we convert to the labe frame time $t$
using $dt={d\tau\over\y}$ EQ.~\ref{eq_lorentzforce} yields (setting $c=1$):
\begary{rcl}
\y\,{d{\cal E}\over dt}&=&q\,\y\,\vec v\cdot\vec E\\
\y\,{d\vec p\over dt}&=&{q\over m}\,\left(m\,\y\,\vec E+m\,\y\,\vec v\times\vec B\right)\\
{d E\over dt}&=&q\,\vec v\cdot\vec E\\
{d\vec p\over dt}&=&q\,\left(\vec E+\vec v\times\vec B\right)\,,
\label{eq_lorentzforce2}
\endary
which is the general form of the Lorentz force. Therefore the Lorentz force 
acting on a charged particle in 3 spatial dimensions can be modeled by an
ensemble of 2-dimensional CHOs. The isomorphism between the observables of the 
perceived 3-dimensional world and the second moments of density distributions 
in the phase space of 2-dimensional oscillators is remarkable.

In any case, Eq.~\ref{eq_lorentzforce} clarifies two things within the game. Firstly,
that both, energy ${\cal E}$ and momentum $\vec p$, have to be interpreted
as mechanical energy and momentum (and not canonical), secondly the relative 
normalization between fields and mechanical momentum is fixed and last, but not least,
it clarifies the relation between the time related to mass (proper time) and the time
related to $\y_0$ and energy, which appears to be the laboratory time.

\section{Collective Motion and the ``Spin''}
\label{sec_Collective}

Eqn.~\ref{eq_envelope} describes the motion of the second moments of
the phase space distribution. Since the equation is linear, it is possible
to bring it into a form $\dot f={\bf B}\,f$ with a ten-component ``vector''
$f$~\cite{rdm_paper}. The frequencies $\Omega_k$ of the modes of collective motion
are given by the ten eigenvalues $\lambda_k=i\,\Omega_k$ of the $10\times 10$ 
matrix ${\bf B}$. Two of these eigenvalues are zero and correspond to a
{\it matched} phase space distribution. The frequencies of the other modes 
are given by
\begary{rcl}
\W_1&=&\pm\sqrt{K1+2\,\sqrt{K2}}\\
\W_2&=&\pm\sqrt{K1-2\,\sqrt{K2}}\\
\W_3&=&\pm{1\over\sqrt{2}}\,\sqrt{K1+\sqrt{K1^2-4\,K2}}\\
\W_4&=&\pm{1\over\sqrt{2}}\,\sqrt{K1-\sqrt{K1^2-4\,K2}}\\
\endary
The frequencies $\W_1$ and $\W_2$ are identical to the frequencies of motion
of a single phase space point (Eqn.~\ref{eq_eigenfreq}). If they are excited,
the averages of the complete distribution oscillate like single spinors.
We might call them coherent modes, as the complete ensemble coherently 
oscillates in these modes, much like the coherent Glauber states or like the 
betatron oscillation of a non-centered beam in a particle accelerator. 
A matched beam (or density distribution) is centered, if all first moments 
of $\psi$ are constantly zero. If we add a fixed centering error $\delta\psi$ 
to a matched $\psi$, then the combined distribution ${\bf S}$ yields:
\begary{rcl}
{\bf\tilde S}&=&\langle(\psi+\delta\psi)\,(\psi+\delta\psi)^T\rangle\,\y_0\\
       &=&\langle\psi\psi^T+\delta\psi\delta\psi^T+\delta\psi\,\psi^T+\psi\,\delta\psi^T\rangle\,\y_0\\
       &=&{\bf S}+(\delta\psi\delta\psi^T)\,\y_0\\
       &=&{\bf S}+\delta{\bf S}\\
\endary
The envelope which can be represented by the second moments oscillates as a
whole about the center. However, the oscillation generates a symplex
$\delta{\bf S}$ with vanishing determinant. As we will show in the next 
section this can be interpreted as the generation of a lightlike spinor.
At the same time, this is a possible Ansatz for the development of a 
``mechanical'' model of how the oscillatory motion of ${\bf S}={\bf P}$ 
generates electromagnetic waves, since we can describe oscillations by 
positions, but also as an oscillation in momentum space. In the
latter case, we may use the proper (absolute time in the comoving frame)
time.

The intrinsic collective modes are those related to the frequencies
$\W_3$ and $\W_4$. There is a high frequency mode with frequency 
$\W_3$ and a low frequency mode$ \pm\W_4$. 
Note further, that in the limit $K_2\to 0$, these modes degenerate: 
the low frequency mode vanishes while the high frequency mode has the
same frequency as the modes of individual motion. However, the collective
modes depend on the same two invariants $K_1$ and $K_2$ as the single modes.
Since these are invariant under (symplectic) transformations, we can describe
collective motion as well in normal coordinates, where the matrix ${\bf F}$
has the form:
\begary{rcl}
{\bf F}&=&\bmtx{cccc}
0&\w_1&0&0\\
-\w_1&0&0&0\\
0&0&0&\w_2\\
0&0&-\w_2&0\emtx\\
&=&{\cal E}\,\y_0+B_y\,\y_8\\
\endary
These are characterized by
$\vec E=\vec P=0$ and $\vec B=B_y\,\vec e_y$. Hence only mass, energy and 
a gyroscopic quantity $\vert\vec B\vert$ are non-zero. Then we have $K_1={\cal E}^2+B_y^2$
and $K_2={\cal E}^2\,B_y^2$, so that the frequencies are:
\begary{rcl}
\W_1&=&{\cal E}+B_y=\w_1\\
\W_2&=&{\cal E}-B_y=\w_2\\
\W_3&=&{\cal E}={\w_1+\w_2\over 2}\\
\W_4&=&B_y={\w_1-\w_2\over 2}\\
\label{eq_normfreq}
\endary
The first two states - the stationary states - show the frequency spectrum
of a spin-$\frac{1}{2}$-particle in an external magnetic field. The last two,
the collective modes, correspond to the average frequency of the two
oscillators and half of the frequency difference, respectively. Hence the
energy of the spin in an external magnetic field is proportional to the frequency
difference between the two oscillators. 

The energy levels - and hence the measurable frequencies $\tilde\w_k$ - of the 
modes described by Eqn.~\ref{eq_normfreq}, are given by:
\begary{rcl}
m^2&=&{\cal E}^2+B_y^2\\
\tilde\w_1&=&\sqrt{m^2-B_y^2}+B_y\\
\tilde\w_2&=&\sqrt{m^2-B_y^2}-B_y\\
\endary
However, up to now the mass $m$ (and the energy) are understood as properties
{\it of the particle} while $B_y$ is an external magnetic field, for which 
$B_y\ll m$\footnote{The magnetic frequency $\w_b={e\over 2\,m}\,B$ equals the
``eigenfrequency'' of the electron $\w_e={m_e\,c^2\over\hbar}$ in case that
$B\approx{10^{10}\,\mathrm{T}}$. Magnetic fields of this size would -
  according to our model - significantly reduce the electron
self-energy - though the energy difference between ``spin up'' and ``spin
down'' remains unchanged.} so that the measurable 
frequencies are simply ${\cal E}\pm B_y$.
 
If the oscillator energy has a contribution that is proportional
to an external field, then we say that the system has a magnetic moment
that is usually interpreted classically by a current loop and hence by an 
intrinsic angular momentum called {\it spin} $\vec S$. In 
section~\ref{sec_LorentzForce} we found the correct scale for the Lorentz 
force, if the fields $\vec E$ and $\vec B$ are scaled by ${q\over 2\,m}$, so that
\begeq
{\cal E}_{magn}=g_e\,{e\over 2\,m_e}\,B_y\,S_y\,,
\endeq
where $g_e$ is the so-called $g$-factor of the electron. Since the two
energy levels are interpreted by the ``spin component'' parallel to the
field ${\bf S}_y=\pm\,{\hbar\over 2}$, we may conclude that $g_e=2$, as
expected. But the idea of a ``charge'' that really ``rotates'' in 
space-time thus generating a magnetic moment is - in view of our 
oscillator model - at best a metaphor.
 
The ``classical'' picture in spacetime requires either a distributed 
charge density in some rotary motion or a point particle in circular 
motion to explain the magnetic moment. The former implies a finite 
radius of the charge density, the latter fails since an 
oscillating charge must - according to electrodynamics - radiate and 
therefore continuously loose energy. Both pictures are 
intrinsically inconsistent while there are no contradictions within
our modeling game with respect to ``spin''. As spacetime is only defined 
by the interaction of several ``objects'' involving a ``distance''. 
In our abstract oscillator model ``spin'' is a ``tune-shift'' between 
coupled oscillators by the ``gyroscopic force'' $B_y$. But the oscillation 
itself is a motion in an abstract phase space - it is a consequence 
of classical oscillator algebra and does not imply any ``real'' motion 
in any ``real'' space-time. 

\subsection{The Maxwell Equations}
\label{sec_maxwell}

As we already pointed out, waves are (within this game) the result of a Fourier 
transformation (FT). But there are different ways to argue this. In Ref.~\cite{qed_paper}
we argued that Maxwell's equations can be derived within our framework by (a) the
postulate that space-time emerges from interaction, i.e. that the 
fields $\vec E$ and $\vec B$ have to be constructed from the 4-vectors.
${\bf X}=t\,\y_0+\vec x\cdot\vec\y$, ${\bf J}=\rho\,\y_0+\vec j\cdot\vec\y$
and ${\bf A}=\Phi\,\y_0+\vec A\cdot\vec\y$ with (b) the requirement that no 
co-symplices emerge. But we can also argue with the FT of the density 
(see Sec.~\ref{sec_density}). 

If we introduce the 4-derivative
\begeq
\d\equiv -\d_t\,\y_0+\d_x\,\y_1+\d_y\,\y_2+\d_z\,\y_3\,.
\label{eq_deriv}
\endeq
The non-abelian nature of matrix multiplication requires to distinguish differential operators 
acting to the right and to the left, i.e. we have $\d$ as defined in Eq.~\ref{eq_deriv}, $\rightD{\d}$ and
$\leftD{\d}$ which is written to the right of the operand (thus indicating the order of the matrix multiplication) 
so that 
\begary{rcl}
{\bf H}\leftD{\d}&\equiv& -\d_t\,{\bf H}\,\y_0+\d_x\,{\bf H}\,\y_1+\d_y\,{\bf H}\,\y_2+\d_z\,{\bf H}\,\y_3\\
\rightD{\d}{\bf H}&\equiv& -\y_0\,\d_t\,{\bf H}+\y_1\,\d_x\,{\bf H}+\y_2\,\d_y\,{\bf H}+\y_3\,\d_z\,{\bf H}\,.
\endary
The we find the following general rules (see Eq.~\ref{eq_cosy_algebra}) that prevent from non-zero cosymplices:
\begary{rcl}
\frac{1}{2}\,\left(\rightD{\d}\textrm{ vector }-\textrm{ vector }\leftD{\d}\right)&\Rightarrow&\textrm{ bi-vector}\\
\frac{1}{2}\,\left(\rightD{\d}\textrm{ bi-vector }-\textrm{ bi-vector }\leftD{\d}\right)&\Rightarrow&\textrm{ vector}\\
\frac{1}{2}\,\left(\rightD{\d}\textrm{ bi-vector }+\textrm{ bi-vector }\leftD{\d}\right)&\Rightarrow&\textrm{ axial vector}=0\\
\frac{1}{2}\,\left(\rightD{\d}\textrm{ vector }+\textrm{ vector }\leftD{\d}\right)&\Rightarrow&\textrm{ scalar}=0\\
\label{eq_difftypes}
\endary
Application of these derivatives yields:
\begary{rcl}
{\bf F}&=&\frac{1}{2}\,\left(\rightD{\d}{\bf A}-{\bf A}\leftD{\d}\right)\\
4\,\pi\,{\bf J}&=&\frac{1}{2}\,\left(\rightD{\d}{\bf F}-{\bf F}\leftD{\d}\right)\\
0&=&\rightD{\d}{\bf F}+{\bf F}\leftD{\d}\\
0&=&\frac{1}{2}\,\left(\rightD{\d}{\bf A}+{\bf A}\leftD{\d}\right)\\
0&=&\frac{1}{2}\,\left(\rightD{\d}{\bf J}+{\bf J}\leftD{\d}\right)\,,
\label{eq_maxwell}
\endary
The first row of Eq.~\ref{eq_maxwell} corresponds to the usual definition of the bi-vector fields
from a vector potential ${\bf A}$ and is (written by components) given by
\begary{rcl}
\vec E&=&-\vec\nabla\phi-\d_t\vec A\\
\vec B&=&\vec\nabla\times\vec A\,.
\endary
The second row of Eq.~\ref{eq_maxwell} corresponds to the usual definition of the 4-current ${\bf J}$
as sources of the fields and the last three rows just express the impossibility of the appearance
of cosymplices. They explicitely represent the homogenuous Maxwell equations
\begary{rcl}
\vec\nabla\cdot\vec B&=&0\\
\vec\nabla\times\vec E+\d_t\vec B&=&0\,,
\label{eq_MWhomo}
\endary
the continuity equation
\begary{rcl}
\d_t\rho+\vec\nabla\vec j=0\,.
\label{eq_continuity}
\endary
and the so-called ``Lorentz gauge''
\begary{rcl}
\d_t\Phi+\vec\nabla\vec A=0\,.
\label{eq_lgauge}
\endary
The simplest idea about the 4-current within QED is to assume that it is
proportional to the ``probability current'', which is within our game given
by the vector components of ${\bf S}=\Sigma\,\y_0$.

\section{The Phase Space}
\label{sec_mass}

Up to now, our modeling game referred to the second moments and the
elements of ${\bf S}$ are second moments such that the observables
are given by (averages over) the following quadratic forms:
\begary{rcl}
{\cal E}&\propto&\psi^T\,\psi=q_1^2+p_1^2+q_2^2+p_2^2\\
p_x&\propto&-q_1^2+p_1^2+q_2^2-p_2^2\\
p_y&\propto& 2\,(q_1\,q_2-p_1\,p_2)\\
p_z&\propto& 2\,(q_1\,p_1+q_2\,p_2)\\
E_x&\propto& 2\,(q_1\,p_1-q_2\,p_2)\\
E_y&\propto&-2\,(q_1\,p_2+q_2\,p_1)\\
E_z&\propto& q_1^2-p_1^2+q_2^2-p_2^2\\
B_x&\propto& 2\,(q_1\,q_2+p_1\,p_2)\\
B_y&\propto& q_1^2+p_1^2-q_2^2-p_2^2\\
B_z&\propto& 2\,(q_1\,p_2-p_1\,q_2)\\
\label{eq_qforms}
\endary
If we analyze the real Dirac matrix 
coefficents of ${\bf S}=\psi\,\psi^T\,\y_0$ in terms of the EMEQ and 
evaluate the quadratic relations between those coefficients, then we 
obtain:
\begary{rcl}
\vec P^2&=&\vec E^2=\vec B^2={\cal E}^2\\
0&=&\vec E^2-\vec B^2\\
{\cal E}^2&=&\frac{1}{2}\,(\vec E^2+\vec B^2)\\
{\cal E}\,\vec P&=&\vec E\times\vec B\\
{\cal E}^3&=&\vec P\cdot(\vec E\times\vec B)\\
m^2&\propto&{\cal E}^2-\vec P^2=0\\
\vec P\cdot\vec E&=&\vec E\cdot\vec B=\vec P\cdot\vec B=0\\
\label{eq_4thorder}
\endary
Besides a missing renormalization these equations describe an 
object without mass but with the geometric properties of light 
as decribed by electrodynamics, e.g. by the electrodynamic description
of electromagnetic waves, which are $\vec E\cdot\vec B=0$, $\vec P\propto\vec
E\times\vec B$, $\vec E^2=\vec B^2$ and so on. Hence single spinors 
are {\it light-like} and can not represent massive particles. 

Consider the spinor as a vector in a four-dimensional Euclidean space.
We write the symmetric matrix ${\cal A}$ (or $\Sigma$, respectively)
as a product in the form of a Gramian:
\begeq
{\cal A}={\cal B}^T\,{\cal B}\,,
\label{eq_spinor1}
\endeq
or - componentwise:
\begary{rcl}
{\cal A}_{ij}&=&\sum_k\,({\cal B}^T)_{ik}\,{\cal B}_{kj}\\
             &=&\sum_k\,{\cal B}_{ki}\,{\cal B}_{kj}\\
\label{eq_spinor2}
\endary
The last line can be read such that matrix element
${\cal A}_{ij}$ is the conventional 4-dimensional scalar 
product of column vector ${\cal B}_i$ with column vector 
${\cal B}_j$. 

From linear algebra we know that Eqn.~\ref{eq_spinor1} yields
a non-singular matrix ${\cal A}$, iff the column-vectors of
the matrix ${\cal B}$ are linearily independent. In the
orthonormal case, the matrix ${\cal A}$ simply is the pure form
of a non-singular matrix, i.e. the unit matrix. Hence, if we
want to construct a massive object from spinors, we need several
spinors to fill the columns of ${\cal B}$. The simplest case is
the orthogonal case: the combination of four mutual orthogonal
vectors. Given a general 4-component Hamiltonian spinor 
$\psi=(q_1,p_1,q_2,p_2)$, how do we find a spinor that is orthogonal 
to this one?
In 3 (i.e. odd) space dimensions, we know that there are two 
vectors that are perpendicular to any vector $(x,y,z)^T$, but
without fixing the first vector, we can't define the others.
In even dimensions this is different: if we fine a non-singular 
skew-symmetric matrix like $\y_0$, then we have a general
expression for a vector that is perpendicular to $\psi$, namely 
$\y_0\,\psi$.
As in Eq.~\ref{eq_heqom}, it is the pure form of the matrix that 
ensures the orthogonality.
Any third vector $\y_k\,\psi$, which is perpendicular to $\psi$ and 
to $\y_0\,\psi$, must then be skew-symmetric {\it and} must fulfil 
$\psi^T\y_k^T\,\y_0\,\psi=0$, which means that the product $\y_k^T\,\y_0$
must be skew-symmetric and hence that $\y_k$ must anti-commute with 
$\y_0$:
\begary{rcl}
(\y_k^T\,\y_0)^T&=&\y_0^T\,\y_k=-\y_k^T\,\y_0\\
\Rightarrow&&\y_0^T\,\y_k+\y_k^T\,\y_0=0\\
0&=&\y_0\,\y_k+\y_k\,\y_0\,.
\endary
Now let us for a moment return to the question of dimensionality.
There are in general $2\,n\,(2\,n-1)/2$ non-zero independent elements 
in a skew-symmetric square $2n\times 2n$ matrix. But how many matrices
are there in the considered phase space dimensions, i.e. in $1+1$, $3+1$ and 
$9+1$ (etc) dimensions which anti-commute with $\y_0$? We need at least 
$2\,n-1$ skew-symmetric anti-commuting elements to obtain a diagonal
${\cal A}$. However, this implies at least $N-1$ anticommuting elements 
of the Clifford algebra that square to $-{\bf 1}$. Hence the ideal case is 
$2n=N$, which is only true for the Pauli and Dirac algebra. For the Pauli
algebra, there is one skew-symmetric element, namely $\eta_0$. 
In the Dirac algebra there are $6$ skew-symmetric generators that 
contain two sets of mutually anti-commuting skew-symmetric matrices:
$\y_0$, $\y_{10}$ and $\y_{14}$ on the one hand and $\y_7$, $\y_8$ and $\y_9$
on the other hand. The next considered Clifford algebra with 
$N=9+1$ dimensions requires a representation by $2n=32=\sqrt{2^{10}}$
-dimensional real matrices. Hence this algebra may not represent a Clifford
algebra with more than $10$ unit elements - certainly not $2\,n$. 
Hence we can not use the algebra to generate purely massive objects 
(e.g. diagonal matrices) without further restrictions (i.e. projections) 
of the spinor $\psi$.

But what exactly does this mean? Of course we can easily find $32$ linearily
independent spinors to generate an orthogonal matrix ${\cal B}$. So what 
exactly is special in the Pauli- and Dirac algebra? To see this, we
need to understand, what it means that we can use the matrix ${\cal B}$ 
of mutually orthogonal column-spinors
\begeq
{\cal B}=(\psi,\y_0\,\psi,\y_{10}\,\psi,\y_{14}\,\psi)\,.
\endeq
This form implies that we can define the {\it mass} of the ``particle''
{\it algebraically}, and since we have $N-1=3$ anticommuting skew-symmetric
matrices in the Dirac algebra, we can find a multispinor ${\cal B}$ for
{\it any} arbitrary point in phase space. This does not seem to be sensational
at first sight, since this appears to be a property of any Euclidean space. 
The importance comes from the fact that $\psi$ is a ``point''
in a very special space - a point in phase space. In fact, we will argue in
the following that this possibility to factorize $\psi$ and the density $\rho$ 
is everything but self-evident.

If we want to simulate a phase space distribution, we can either define a 
phase space density $\rho(\psi)$ or we use the
technique of Monte-Carlo simulations and represent the phase space
by (a huge number of random) samples. If we generate a random sample 
and we like to implement a certain exact symmetry of the density in phase 
space, then we would (for instance) form a symmetric sample by appending not 
only a column-vector to ${\cal B}$, but also its negative $-\psi$. In this way
we obtain a sample with an exact symmetry.
In a more general sense: If a phase space symmetry can be represented by a 
matrix $\y_s$ that allows to associate to an arbitrary phase space point 
$\psi$ a second point $\y_s\,\psi$ where $\y_s$ is skew-symmetric, then we 
have a certain continuous linear rotational symmetry in this phase space.
As we have shown, phase-spaces are intrinsically structured by $\y_0$ and
insofar much more restricted than Euclidean spaces. This is due to the 
distinction of symplectic from non-symplectic transformations and due 
to the intrinsic relation to Clifford algebras:
{\it Phase spaces are spaces structured by time}. According to the rules 
and results of the game, they appear to be the only physically fundamental 
spaces at all.

We may imprint the mentioned symmetry to an arbitrary 
phase space density $\rho$ by taking all phase space samples that
we have so far and adding the same number of samples, each column
multiplied by $\y_s$. Thus, we have a single rotation in the
Pauli algebra and two of them in the Dirac algebra:
\begary{rcl}
&&{\cal B}_0=\psi\\
\y_0&\to&{\cal B}_1=(\psi,\y_0\,\psi)\\
\y_{14}&\to&{\cal B}_2=(\psi,\y_0\,\psi,\y_{14}\,\psi,\y_{14}\,\y_0\,\psi)\\
&&=(\psi,\y_0\,\psi,\y_{14}\,\psi,\y_{10}\,\psi)\\
\label{eq_ft1}
\endary
or: 
\begary{rcl}
&&{\cal B}_0=\psi\\
\y_7&\to&{\cal B}_1=(\psi,\y_7\,\psi)\\
\y_8&\to&{\cal B}_2=(\psi,\y_7\,\psi,\y_8\,\psi,\y_8\,\y_7\,\psi)\\
&&=(\psi,\y_7\,\psi,\y_8\,\psi,-\y_9\,\psi)\\
\label{eq_ft2}
\endary
Note that order and sign of the column-vectors in ${\cal B}$ are irrelevant 
- at least with respect to the autocorrelation matrix ${\cal B}\,{\cal B}^T$.
Thus we find that there are two fundamental ways to represent a positive
mass in the Dirac algebra and one in the Pauli-algebra. The $4$-dimensional
phase space of the Dirac algebra is in two independent ways self-matched.

Our starting point was the statement that $2\,n$ linear independent
vectors are needed to generate mass. If we can't find $2\,n$ vectors
in the way described above for the Pauli and Dirac algebra, then this 
does (of course) not automatically imply that there are not $2\,n$ linear 
independent vectors.  

But what does it mean that the dimension of the Clifford algebra of
observables ($N$) does not match the dimension of the phase space ($2\,n$)
in higher dimensions? There are different physical descriptions given. 
Classically we would say that a positive definite $2\,n$-component spinor 
describes a system of $n$ (potentially) coupled oscillators with $n$ frequencies.
If ${\cal B}$ is orthogonal, then all oscillators have the same frequency,
i.e. the system is degenerate. But for $n>2$ we find that not all eigenmodes 
can involve the complete $2\,n$-dimensional phase space. This phenomenon
is already known in $3$ dimensions: The trajectory of the isotropic 
three-dimensional oscillator always happens in a $2$-dimensional plane,
i.e. in a subspace. If it did not, then the angular momentum would not
be conserved. In this case the isotropy of space would be broken.
Hence one may say in some sense that the {\it isotropy of space} is the
reason for a $4$-dimensional phase-space and hence the reason for
the $3+1$-dimensional observable space-time of objects. Or in other
words: higher-dimensional spaces are incompatible with isotropy, i.e.
with the conservation of angular momentum. There is an intimate connection
of these findings to the impossibility of Clifford algebras
$Cl_{p,1}$ with $p>3$ to create a homogeneous ``Euclidean'' space: Let
$\y_0$ represent time and $\y_k$ with $k\in[1,\dots,N-1]$ the spatial
coordinates. The spatial rotators are products of two spatial basis
vectors. The generator of rotations in the $(1,2)$-plane is $\y_1\,\y_2$.
Then we have $6$ rotators in $4$ spatial dimensions:
\begary{cccccc}
\y_1\,\y_2,&\y_1\,\y_3,&\y_1\,\y_4,&\y_2\,\y_3,&\y_2\,\y_4,&\y_3\,\y_4\,.
\endary
However, we find that some generators commute and while others anticommute and
it can be taken from combinatorics that only sets of $3$ mutual anti-commuting
rotators can be formed from a set of symmetric anti-commuting $\y_k$. 
The $3$ rotators 
\begary{ccc}
\y_1\,\y_2,&\y_2\,\y_3,&\y_1\,\y_3\\
\endary
mutually anticommute, but $\y_1\,\y_2$ and $\y_3\,\y_4$ commute.
Furthermore, in $9+1$ dimensions, the spinors are either projections
into 4-dimensional subspaces or there are non-zero off-diagonal terms 
in ${\cal A}$, i.e. there is ``internal interaction''.

Another way to express the above considerations is the following: Only
in $4$ phase space dimensions we may construct a massive object from
a matrix ${\cal B}$ that represents a multispinor $\Psi$
of exactly $N=2n$ single spinors and construct a wave-function according to
\begeq
\Psi=\phi\,{\cal B}\,,
\label{eq_multispinor0}
\endeq
where $\rho=\phi^2$ is the phase space density.

It is easy to prove and has been shown in Ref.~\cite{qed_paper} that 
the elements $\y_0$, $\y_{10}$ and $\y_{14}$ represent parity, time
reversal and charge conjugation. The combination of these operators 
to form a multispinor, may lead (with normalization) to the construction
of symplectic matrices ${\bf M}$. Some examples are:
\begary{rcl}
{\bf M}&=&({\bf 1}\,\psi,\y_0\,\psi,-\y_{14}\,\psi,-\y_{10}\,\psi)/\sqrt{\psi^T\,\psi}\\
{\bf M}\,\y_0\,{\bf M}^T&=&\y_0\\
&&\\
{\bf M}&=&({\bf 1}\,\psi,-\y_{14}\,\psi,-\y_{10}\,\psi,\y_0\,\psi)/\sqrt{\psi^T\,\psi}\\
{\bf M}\,\y_{10}\,{\bf M}^T&=&\y_{10}\\
&&\\
{\bf M}&=&(\y_{10}\,\psi,-{\bf 1}\,\psi,-\y_{14}\,\psi,\y_0\,\psi)/\sqrt{\psi^T\,\psi}\\
{\bf M}\,\y_{14}\,{\bf M}^T&=&\y_{14}\\
\endary
Hence the combination of the identity and CPT-operators can be arranged such
that the multispinor ${\bf M}$ is symplectic with respect to the directions 
of time $\y_0$, $\y_{10}$ and $\y_{14}$, but not with respect to $\y_7$,
$\y_8$ or $\y_9$. As we tried to explain, the specific choice of the 
skew-symmetric matrix $\y_0$ is determined by a structure defining
transformation. Since particles are nothing but dynamical structures in this
game, the $6$ possible SUMs should stand for $6$ different particle types.
However, for each direction of time, there are also two choices of the 
spatial axes. For $\y_0$ we have chosen $\y_1$, $\y_2$ and $\y_3$, but
we could have used $\y_4=\y_0\,\y_1$, $\y_5=\y_0\,\y_2$ and $\y_6=\y_0\,\y_3$
as well.

Thus, there should be either 6 or 12 different types of structures (types of fermions)
that can be constructed within the Dirac algebra. The above construction allows for
$3$ different types corresponding to $3$ different forms of the symplectic unit matrix,
further $3$ types are expected to be related to $\y_7$, $\y_8$ and $\y_9$:
\begary{rcl}
{\bf M}&=&({\bf 1}\,\psi,-\y_9\,\psi,-\y_{8}\,\psi,-\y_{7}\,\psi)/\sqrt{\psi^T\,\psi}\\
{\bf M}\,\y_7\,{\bf M}^T&=&\y_7\\
&&\\
{\bf M}&=&({\bf 1}\,\psi,-\y_{8}\,\psi,-\y_{7}\,\psi,-\y_9\,\psi)/\sqrt{\psi^T\,\psi}\\
{\bf M}\,\y_{8}\,{\bf M}^T&=&\y_{8}\\
&&\\
{\bf M}&=&(\y_{7}\,\psi,-{\bf 1}\,\psi,-\y_{8}\,\psi,-\y_9\,\psi)/\sqrt{\psi^T\,\psi}\\
{\bf M}\,\y_{9}\,{\bf M}^T&=&\y_{9}\\
\endary
These matrices describe specific symmetries of the 4-dimensional phase space, i.e. geometric
objects in phase space. Therefore massive multispinors can be described as volumes in phase space.
If we deform the figure by stretching parameters $a,b,c,d$ such that
\begeq
{\bf\tilde M}=(a\,{\bf 1}\,\psi,-b\,\y_0\,\psi,-c\,\y_{14}\,\psi,-d\,\y_{10}\,\psi)/\sqrt{\psi^T\,\psi}\,,
\endeq
then one obtains with $f_k$ taken from Eq.~\ref{eq_qforms}:
\begary{rcl}
{\bf\tilde M}\,{\bf\tilde M}^T\,\y_0&=&\sum\limits_{k=0}^9\,g_k\,f_k\,\y_k/\sqrt{\psi^T\,\psi}\\
g_0&=& a^2+b^2+c^2+d^2\\
g_1&=&-g_2=g_3=a^2-b^2+c^2-d^2\\
g_4&=&-g_5=g_6=a^2-b^2-c^2+d^2\\
g_7&=&g_8=g_9=a^2+b^2-c^2-d^2\\
\label{eq_phs1}
\endary
This result reproduces the quadratic forms $f_k$ of Eq.~\ref{eq_qforms}, but
furthermore the phase space radii $a$, $b$, $c$ and $d$ reproduce the structure
of the Clifford algebra, i.e. the classification into the 4 types of observables
${\cal E}$, $\vec P$, $\vec E$ and $\vec B$. This means that a deformation 
of the phase space ``unit cell'' represents momenta and fields, i.e. the dimensions 
of the phase space unit cell are related to the appearance of certain symplices:
\begary{rcl}
(a=b)\,\mathrm{AND}\, (c=d)&\Rightarrow&\vec P=\vec E=0\\
(a=c)\,\mathrm{AND}\, (b=d)&\Rightarrow&\vec E=\vec B=0\\
(a=d)\,\mathrm{AND}\, (b=c)&\Rightarrow&\vec P=\vec B=0\,,
\label{eq_phs2}
\endary
while for $a=b=c=d$ all vectors but ${\cal E}$ vanish. Only in this latter
case, the matrix ${\bf M}$ is symplectic for $a=b=c=d=1$. These relations
confirm the intrinsic connection between a classical 4-dimensional
Hamiltonian phase space and Clifford algebras in dimension 3+1.

\section{Summary and Discussion}
\label{sec_summary}

Based on three fundamental principles we have shown that the algebraic
structure of coupled classical degrees of freedom is (depending on the
number of the DOFs) isomorph to certain Clifford algebras that allow  
to explain the dimensionality of space-time, to model Lorentz-transformations, 
the relativistic energy-momentum relations and even Maxwell's equations.

It is usually assumed that we have to define the properties of space-time in the first
place: ``In Einstein's theory of gravitation matter and its dynamical interaction 
are based on the notion of an intrinsic geometric structure of the space-time continuum''\cite{ES1950}.
However - as we have shown within this ``game'' - it has far more explanatory power 
to derive and explain space-time from the principles of interaction. 
Hence we propose to reverse the above statement: The intrinsic geometric 
structure of the space-time continuum is based on the dynamical interaction 
of matter. A rigorous consequence of this reversal of perspective is that 
``space-time'' does not need to have a fixed and unique 
dimensionality at all. It appears that this dimensionality is {\it relative} to
the type of interaction. However, supposed higher-dimensional space-times
(see Ref.~\cite{qed_paper}) would emerge in analogy to the method presented here,
for instance in nuclear interaction, then these space-times would not simply be
Euclidean spaces of higher dimension. Clifford algebras - especially if they
are restricted by symplectic conditions by a Hamiltonian function, have a
surprisingly complicated intrinsic structure. As we pointed out, if all 
generators of a Clifford algebra are symplices, then in $9+1$ dimensions, 
we find $k$-vectors with $k\in\,[0..10]$ but $k$-vectors generated from
symplices are themselves symplices only for $k\in\,[1,2,5,6,9,10,\dots]$.
However, if space-time is constraint by Hamiltonian motion, then ensembles 
of oscillators may also clump together to form
``objects'' with $9+1$ or $25+1$-dimensional interactions, despite the fact
that we gave strong arguments for the fundamentality of $3+1$-dimensional
Hamiltonian algebra.

There is no a priori reason to exclude higher order terms - whenever they 
include constants of motion. However, as the Hamitonian then involves terms 
of higher order, we might then need to consider higher order moments 
of the phase space distribution. In this case we would have to invent
an action constant in order to scale $\psi$. 

Our game is based a few general rules and symmetry considerations.
The math used in our derivation - taken the results of representation theory 
for granted - is simple and can be understood on an undergraduate level.
And though we never intended to find a connection to string theory, we
found - besides the $3+1$ - dimensional interactions a list of possible 
higher-dimensional candidates, two of which are also in the focus
of string theories, namely $9+1=10$-dimensional and $25+1=26$-dimensional 
theories~\cite{StringTheory}. 

We understand this modeling game as a contribution to the demystification (and
unification) of our understanding of space-time, relativity, electrodynamics
and quantum mechanics. Despite the fact that it has become tradition to write
all equations of motion of QED and QM in a way that requires the use of the unit
imaginary, our model seems to indicate that it does not have to be that way.
Though it is frequently postulated that evolution in time has to be unitary 
within QM, it appears that symplectic motion does not only suffice, but is 
superior as it yields the correct number of relevant operators. 
While in the unitary case, one should expect $16$ ($15$) unitary (traceless) 
operators for a $4$-component spinor, while the natural number of generators in 
the corresponding symplectic treatment is $10$ as found by Dirac himself in 
QED~\cite{Dirac49,Dirac63}.
If a theory contains things which are {\it not required}, then we have
added something arbitrary and artificial. The theory as we described it 
indicates that in momentum space, which is used here, there is 
no immediate need for the use of the unit imaginary and no need for more 
than $10$ fundamental generators. The use of the unit imaginary however
appears unavoidable when we switch via Fourier transform to the ``real space''. 

There is a dichotomy in physics. On the one hand all {\it causes} are 
considered to inhabit space-time ({\it local causality}), but on the other 
hand the {\it physical reasoning} mostly happens in energy- or momentum space: 
There are no Feyman-graphs, no scattering amplitudes, no fundamental 
physical relations, that do not refer in some way to energy or momentum 
(-conservation). We treat problems in solid state physics as well as
in high energy physics mostly in Fourier space (reciprocal lattice).

We are aware that the rules of the game are with their rigour difficult to 
accept. However, maybe it does not suffice to speculate that the world might 
be a hologram~\footnote{As t'Hooft suggested~\cite{tHooft} and Leonard
  Susskind sketched in his celebrated paper, Ref.~\cite{Susskind}.
} - we really should play modeling games that might help to decide, if and
{\it how} it could be like that.

\begin{appendix}

\section{Microcanonical Ensemble}
\label{sec_micro}

Einstein once wrote that ``A theory is the more impressive the greater
the simplicity of its premises, the more different kinds of things it relates,
and the more extended its area of applicability. Hence the deep impression
that classical thermodynamics made upon me. It is the only physical theory
of universal content concerning which I am convinced that, within the
framework of the applicability of its basic concepts, it will never be
overthrown [...]''~\cite{EinsteinBio}. We agree with him and we will try
to show in the following that this holds also for the branch of thermodynamics
that is called statistical mechanics. 

By the use of the EMEQ it has been shown, that the expectation values 
\begeq
f_k={\mathrm{Tr}(\y_k^2)\over 16}\,\bar\psi\,\y_k\,\psi
\endeq
can be associated with energy ${\cal E}$ and momentum $\vec p$ {\it of} and
with the electric (magnetic) field $\vec E$ and $\vec B$ as {\it seen by} a 
relativistic charged particle. It has also been shown that stable systems 
can always be transformed in such a way as to bring ${\cal H}$ into a diagonal form:
\begeq
{\bf F}=\bmtx{cccc}
0&\w_1&0&0\\
-\w_1&0&0&0\\
0&0&0&\w_2\\
0&0&-\w_2&0\emtx\,,
\label{eq_f_reg}
\endeq
In the following we will use the classical model of the 
microcanonical ensemble to compute some phase space averages.
Let the constant value of the Hamiltonian be ${\cal H}=U$ where $U$ 
is some energy, the volume in phase space $\Phi^\star$ that is
limited by the surface of constant energy $U$ is given by~\cite{Becker}:
\begeq
\Phi^\star=\int\limits_{{\cal H}<U}\,dq_1\,dp_1\,dq_2\,dp_2\,,
\label{eq_phstar0}
\endeq
and the partition function $\w^\star$ is the derivative 
\begeq
\w^\star={d\Phi^\star\over dU}\,,
\label{eq_wstar0}
\endeq
which is the phase space integral over all states of constant energy $U$.
The average value of any phase space function $\overline{f(p,q)}$ is then 
given by
\begeq
\overline{f(p,q)}={1\over\w^\star}{d\over dU}\int\limits_{{\cal H}<U}\,f(p,q)\,dq_1\,dp_1\,dq_2\,dp_2\,.
\label{eq_mce1}
\endeq
In case of a 2-dimensional harmonic oscillator, for instance, we
may take the following parametrization of the phase space:
\begary{rcl}
q_1&=&r\,\cos{(\alpha)}\,\cos{(\beta)}\\
p_1&=&r\,\cos{(\alpha)}\,\sin{(\beta)}\\
q_2&=&r\,\sin{(\alpha)}\,\cos{(\y)}\\
p_2&=&r\,\sin{(\alpha)}\,\sin{(\y)}\\
\label{eq_par2}
\endary
Note that Eqn.~\ref{eq_par2} describes a solution of the equations of motion
\begeq
\dot\psi={\bf F}\,\psi
\endeq
when we replace 
\begary{rcl}
\beta&\to&-\w_1\,t\\
\y&\to&-\w_2\,t\,.
\endary
This means that the (normalized) integration over $\beta$ and $\y$ is mathematically
identical to an integration over all times (time average).
From Eqn.~(\ref{eq_mce1}) one would directly conclude
\begeq
\overline{f(p,q)}={1\over\w^\star}{d\over dU}\int\limits_{{\cal H}<U}\,f(p,q)\,\sqrt{g}\,dr\,d\alpha\,d\beta\,d\y\,,
\label{eq_mce2a}
\endeq
where $g$ is the Gramian determinant. However the relative amplitude controlled by
the parameter $\alpha$ can not be changed by symplectic transformations and hence 
remains constant in a closed system. Therefore the phase space trajectory of the 
oscillator can not cover the complete 3-dim. energy surface, but only a 2-dim. subset 
thereof. This is known very well in accelerator physics as the emittance preservation 
of decoupled DOF.
And we have shown in Ref.~\cite{geo_paper} that all stable harmonic oscillators systems
are symplectically similar to a decoupled system. 
Consequently $\alpha$ has to be excluded from the integration of a ``single particle'' 
average and has to be treated instead as an additional parameter or ``boundary condition'':
\begeq
\overline{f(p,q)}={1\over\w^\star}{d\over dU}\int\limits_{{\cal H}<U}\,f(p,q)\,\sqrt{g}\,dr\,d\beta\,d\y\,,
\label{eq_mce2}
\endeq
The Gramian determinant hence is given by:
\begary{rcl}
g&=&\mathrm{Det}({\bf G}^T\,{\bf G})\\
{\bf G}&=&\bmtx{ccc}
{\d q_1\over\d r}&{\d q_1\over\d\beta}&{\d q_1\over\d\y}\\
{\d p_1\over\d r}&{\d p_1\over\d\beta}&{\d p_1\over\d\y}\\
{\d q_2\over\d r}&{\d q_2\over\d\beta}&{\d q_2\over\d\y}\\
{\d p_2\over\d r}&{\d p_2\over\d\beta}&{\d p_2\over\d\y}\\
\emtx\,,
\endary
so that one finds
\begeq
\sqrt{g}=r^2\,\cos{(\alpha)}\,\sin{(\alpha)}\,.
\endeq
Accordingly, Eqn.~(\ref{eq_phstar0}) has to be written as
\begeq
\Phi^\star=\int\limits_{H<U}\,\sqrt{g}\,dr\,d\beta\,d\y\,.
\endeq
We use the abbreviations
\begary{rcl}
{\bar\w}&=&{\w_1+\w_2\over 2}\\
{\Delta\w}&=&{\w_1-\w_2\over 2}\\
\Omega&=&\bar\w+\Delta\w\,\cos{(2\,\alpha)}\,.
\endary
The Hamilton function is given in the new coordinates by
\begeq
{\cal H}={r^2\,\W\over 2}\,,
\label{eq_Hnew1}
\endeq
so that the condition ${\cal H}<U$ translates into
\begeq
r \le \sqrt{2\,\eps}\,,
\endeq
where $\eps={U/\W}$.
The integration over $\beta$ and $\y$ is taken from $0$ to $2\,\pi$.
The complete integration results in
\begary{rcl}
\Phi^\star&=&{2\,\pi^2\over 3}\,\sin{(2\,\alpha)}\,\left({2\,U\over\W}\right)^{3/2}\\
\w^\star&=&\frac{3}{2\,U}\,\Phi^\star\\
\endary
The following average values are computed from Eqn.~(\ref{eq_mce2}):
\begary{rcl}
\overline{\cal H}&=&U\\
\overline{{\cal H}^2}&=&U^2\\
\overline{f_k}&=&\left\{\begin{array}{lcl}
\eps &\mathrm{for}&k=0\\
\eps\,\cos{(2\,\alpha)}&\mathrm{for}&k=8\\
0&\mathrm{for}&k\in \{1-7,9\}\\
\end{array}\right.\\
\overline{f_k^2}&=&\left\{\begin{array}{lcl}
\overline{f_k}^2&\mathrm{for}&k=0,8\\
\frac{1}{4}\,\eps^2\,(1+\cos^2{(2\,\alpha)})&\mathrm{for}&k\in \{1,3,4,6\}\\
\frac{1}{2}\,\eps^2\,\sin^2{(2\,\alpha)}    &\mathrm{for}&k\in \{2,5,7,9\}\\
\end{array}\right.\,.
\label{eq_mcexp}
\endary
Hence we find that $f_0$ (energy), $f_8$ (one spin component) and ${\cal H}$ (mass)
are ``sharp'' (i.e. operators with an eigenvalue), while the other ``expectation values'' 
have a non-vanishing variance. The fact that spin always has a ``direction of quantization'', 
i.e. that only {\it one single} ``sharp'' component, can therefore be nicely modelled within our
game. It is a consequence of symplectic motion.
Note also that the squared expectation values of all even ($\y_1$, $\y_3$, $\y_4$ and $\y_6$, 
except $\y_0$ and $\y_8$) and all odd ($\y_2$, $\y_5$, $\y_7$ and $\y_9$) operators are equal~\footnote{
The {\it even} Dirac matrices are block-diagonal, the {\it odd} ones not. There are six even
symplices and four odd ($\y_2$, $\y_5$, $\y_7$ and $\y_9$) ones~\cite{rdm_paper}. Obviously this 
pattern is the reason for the grouping in Eq.~\ref{eq_mcexp}.
}.

Consider the coordinates are given by the fields ($\vec E\propto\vec x$) $\vec Q=(f_4,f_5,f_6)^T$
and the momenta as usual by $\vec P=(f_1,f_2,f_3)^T$, then the angular momentum $\vec L$ should
be given by $\vec L=\vec Q\times\vec P$. We obtain
the following expectation values from the microcanonical ensemble:
\begary{rcl}
\overline{L_x}&=&\overline{L_z}=0\\
\overline{L_x^2}&=&\overline{L_z^2}=\frac{1}{2}\,\eps^4\,\sin^2{(2\,\alpha)}\\
\overline{L_y}&=&\eps^2\,\cos{(2\,\alpha)}\\
\overline{L_y^2}&=&\eps^4\,\cos^2{(2\,\alpha)}\\
\endary
That is - up to a common scale factor of $\eps$ (or $\eps^2$, respectively) - we have the same 
results as in Eq.~\ref{eq_mcexp}. 
Consider now the quantum mechanical postulates that the spin component of a fermion
is $s_z=\pm s=\pm\frac{1}{2}$ and $\vert\vec s\vert^2=s_x^2+s_y^2+s_z^2=s(s+1)=\frac{3}{4}$.
We can ``derive'' this result (up to a factor) from an isotropy requirement for the 4th order
moments, i.e. from the condition that $\langle P_x^2\rangle=\langle P_y^2\rangle=\langle P_z^2\rangle$:
\begeq
\sin^2{(2\,\alpha)}=2\,\cos^2{(2\,\alpha)}
\endeq
so that
\begeq
\alpha=\frac{1}{2}\,\arctan{\sqrt{2}}=27.3678^\circ\,,
\endeq
or equivalently with
\begeq
1+3\,\cos^2{(2\,\alpha)}=0
\endeq
we obtain
\begeq
\cos{(2\,\alpha)}=\pm\frac{1}{\sqrt{3}}\,.
\endeq
With respect to the symplex ${\bf F}$ as defined in Eq.~\ref{eq_f_reg}, we have
\begeq
{\bf F}=\bar\w\,\y_0+\Delta\w\,\y_8\,,
\endeq
so that with Eq.~\ref{eq_mcexp} one finds
\begeq
{f_8/f_0}={\overline{f_8}/\overline{f_0}}=\cos{(2\,\alpha)}
\label{eq_freq_ratio}
\endeq
The total spin would then be given by the 4th order moments as
$\vec S^2=\eps^2$, so that for a spin-$\frac{1}{2}$-particle we would
have to normalize to $\eps^2=(s(s+1))\hbar^2=\frac{3}{4}\,\hbar^2$ and
hence $f_8=\sqrt{3\over 4}\,\hbar\,{1\over\sqrt{3}}={\hbar\over 2}$.
However, the mass formula (Eq.~\ref{eq_eigenfreq}), refers to the second moments, 
so that in linear theory we would have
\begeq
{\cal H}=m=\omega_0\,\sqrt{f_0^2+f_8^2}=\omega_0\,\sqrt{\eps^2+\frac{1}{3}\,\eps^2}=\omega_0\,\sqrt{\frac{4}{3}}\,\eps=\omega_0\,\hbar
\endeq
In order to relate this to a frequency difference, we use Eq.~\ref{eq_freq_ratio}:
\begary{rcl}
{f_8\over f_0}&=&{\Delta\w\over\bar\w}={\w_1-\w_2\over\w_1+\w_2}=\cos{(2\,\alpha)}=\frac{1}{\sqrt{3}}\\
\Rightarrow&&\\
{\w_1\over\w_2}&=&2+\sqrt{3}\\
\W&=&4\,\Delta\w=\frac{4}{\sqrt{3}}\,\bar\w\,.
\endary
Then from $r=\sqrt{2\,\eps}$ and Eq.~\ref{eq_Hnew1} we find
\begary{rcl}
{\cal H}&=&{r^2\,\W\over 2}=\eps\,\W=\sqrt{3\over4}\,\hbar\,\W=\sqrt{3\over4}\,\hbar\,\frac{4}{\sqrt{3}}\,\bar\w\\
        &=&2\,\hbar\,\bar\w=\hbar (\w_1+\w_2)\,.
\endary
To conclude, classical statistical mechanics allows for a description of 
spin, if the rules of symplectic motion are taken into account. 
This alone is remarkable. Secondly, assumed that the microcanonical ensemble is the right
approach, then the isotropy of the emergent 3+1-dimensional space-time (with respect to 
4th-order moments) apparently requires a certain ratio between the frequencies and 
amplitudes of the two coupled oscillators, i.e. an asymmetry on the fundamental level.

\subsection{Entropy and Heat Capacity}

The entropy ${\cal S}$ of the microcanonical ensemble can be written as~\cite{Becker}:
\begeq
{\cal S}=k\,\log{\Phi^\star}\,.
\endeq
The temperature $T$ of the system is given by
\begary{rcl}
{\d{\cal S}\over\d U}&=&{1\over T}={\d (k\,\log{\Phi^\star})\over\d U}=k\,{\omega^\star\over\Phi^\star}\\
                     &=&{3\,k\over 2\,U}\\
\endary
so that energy as a function of temperature is
\begeq
U={3\over 2}\,k\,T\,,
\endeq
and the heat capacity $C_V={\d U\over\d T}$ is (per particle)
\begeq
C_V={3\over 2}\,k\,.
\endeq
This important result demonstrates -- according to statistical mechanics -- 
the 3-dimensionality of the ``particle'' as the energy per DOF is 
${k\,T\over 2}$: a two-dimensional harmonic oscillator of fundamental variables
is equivalent to an free 3-dimensional ``point particle''. To our knowledge this 
is the first {\it real physical} model of a relativistic point particle.

\end{appendix}

\end{document}